\documentclass[journal,comsoc]{IEEEtran}
\usepackage[T1]{fontenc}

\usepackage{array}

\usepackage{graphicx} 
\graphicspath{{./figs/}}

\usepackage{subcaption}
\usepackage{amsmath, amssymb, amsfonts}
\usepackage{color}
\usepackage{mathtools}
\usepackage{algorithm}
\usepackage{algorithmic}
\usepackage{multirow}
\usepackage{float}
\usepackage{url} 
\usepackage{comment}
\usepackage{cite}

\usepackage{cancel} 
\usepackage[normalem]{ulem} 

\newcommand{\stkout}[1]{%
  \ifmmode
    \cancel{#1}%
  \else
    \sout{#1}%
  \fi
}

\newcommand{\add}[1]{{\color{black}#1}}

\usepackage[acronym]{glossaries}
\newacronym{tx}{Tx}{transmitter}
\newacronym{rx}{Rx}{receiver}
\newacronym{mimo}{MIMO}{Multiple-Input Multiple-Output}
\newacronym{mumimo}{MU-MIMO}{Multi-User MIMO}
\newacronym{ula}{ULA}{Uniform Linear Array}
\newacronym{fista}{FISTA}{Fast Iterative Shrinkage-Thresholding Algorithm}
\newacronym{pg}{PG}{projected gradient}
\newacronym{apg}{APG}{accelerated PG}
\newacronym{am}{AM}{alternating minimization}
\newacronym{rcs}{RCS}{radar cross-section}
\newacronym{isac}{ISAC}{Integrated Sensing and Communication}
\newacronym{psl}{PSL}{peak sidelobe level}
\newacronym{isl}{ISL}{integrated sidelobe level}

\newacronym{bpm}{BPM}{beampattern matching}
\newacronym{svd}{SVD}{singular value decomposition}

\newacronym{snr}{SNR}{signal-to-noise ratio}
\newacronym{sinr}{SINR}{signal-to-interference-plus-noise ratio}

\newacronym{mui}{MUI}{multi-user interference}
\newacronym{pa}{PA}{power amplifier}
\newacronym{rf}{RF}{radio-frequency}
\newacronym{par}{PAPR}{peak-to-average-power-ratio}
\newacronym{cm}{CM}{constant-modulus}
\newacronym{qpsk}{QPSK}{quadrature phase shift keying}

\newacronym{mse}{MSE}{mean square error}
\newacronym{wmse}{MSE}{weighted mean square error}

\newacronym{ls}{LS}{least squares}
\newacronym{wls}{WLS}{weighted least squares}

\newacronym{mmse}{MMSE}{minimum mean square error}
\newacronym{zf}{ZF}{zero-forcing}
\newacronym{ser}{SER}{symbol error rate}
\newacronym{mf}{MF}{matched filter}

\newacronym{mrt}{MRT}{maximum ratio transmission}

\newacronym{bs}{BS}{base station}
\newacronym{los}{LoS}{line-of-sight}
\newacronym{dac}{DAC}{digital-to-analog converter}
\newacronym{sdp}{SDP}{semidefinite programming problem}

\newacronym{qsdp}{QSDP}{quadratic semidefinite programming problem}
\newacronym{wf}{WF}{Wiener filter}
\newacronym{csi}{CSI}{channel state information}
\newacronym{qam}{QAM}{quadrature amplitude modulation}
\newacronym{doa}{DOA}{direction-of-arrival}
\newacronym{crb}{CRB}{Cramer-Rao Bound}
\newacronym{lfm}{LFM}{linear frequency modulation}
\newacronym{sdr}{SDR}{semi-definite relaxation}

\newacronym{ismr}{ISMR}{integrated sidelobe to mainlobe ratio}

\newacronym{tdd}{TDD}{time-division-duplex}
\newacronym{dft}{DFT}{discrete Fourier transform}
\newacronym{apes}{APES}{amplitude and phase estimation}
\newacronym{sir}{SIR}{signal-to-interference ratio}
\newacronym{is}{IS}{interference suppression}

\newcommand{\argmin}{\arg\, \min}

\newcommand\figWidth{6} 

\newlength\myindent
\newcommand\bindent{%
	\begingroup
	\setlength{\itemindent}{\myindent}
	\addtolength{\algorithmicindent}{\myindent}
}
\newcommand\eindent{\endgroup}

\newlength\myindentt
\newcommand\bindentt{%
	\begingroup
	\setlength{\itemindent}{\myindentt}
	\addtolength{\algorithmicindent}{\myindentt}
}
\newcommand\eindentt{\endgroup}

\usepackage{tikz}

\newcommand\copyrighttext{%
  \footnotesize \textcopyright \the\year{} IEEE. Personal use of this material is permitted.  Permission from IEEE must be obtained for all other uses, in any current or future media, including reprinting/republishing this material for advertising or promotional purposes, creating new collective works, for resale or redistribution to servers or lists, or reuse of any copyrighted component of this work in other works.}

\newcommand\copyrightnotice{%
\begin{tikzpicture}[remember picture,overlay]
\node[anchor=south,yshift=1pt] at (current page.south) {\fbox{\parbox{\dimexpr0.90\textwidth-\fboxsep-\fboxrule\relax}{\copyrighttext}}};
\end{tikzpicture}%
}

\begin{document}

\title{Waveform Design for Simultaneous MIMO Radar Sensing and Multi-User Communication}
%
%
%
\author{Berkan Kilic,
        Kenan Turbic,~\IEEEmembership{Member,~IEEE,}
        and
        S\l{}awomir Sta\'nczak,~\IEEEmembership{Senior Member,~IEEE}%
\thanks{
The authors acknowledge the financial support by the Federal Ministry for Research, Techonology and Space (BMFTR) in Germany in the programme of ``Souverän. Digital. Vernetzt.'' Joint project 6G-RIC, project identification numbers: 16KISK020K and 16KISK030. (\textit{Corresponding author: {Berkan Kilic}.})

The authors are with the Wireless Communications and Networks Department, Fraunhofer Institute for Telecommunications, Heinrich Hertz Institute (HHI), 10587, Berlin, Germany (e-mail: \{berkan.kilic, kenan.turbic, slawomir.stanczak\}@hhi.fraunhofer.de).}%
\thanks{S. Sta\'nczak is also with Technische Universit{\"a}t Berlin, 10587, Berlin, Germany.}%
}


%



\markboth{}%
{Kilic et al.: Waveform Design for Simultaneous MIMO Radar Sensing and Multi-User Communication}
\maketitle

\copyrightnotice
\begin{abstract}
This paper proposes a novel two-stage joint waveform design framework for multi-antenna \gls{isac} systems that simultaneously enable \gls{mimo} radar sensing and Multi-User \gls{mimo} communication.
First, a transmit waveform covariance matrix is designed by solving a convex matrix nearness problem for beampattern synthesis that simultaneously maximizes transmit power in desired directions and minimizes cross-directional correlations, while accommodating independent antenna power constraints and supporting interference suppression through radiation null steering.
Second, a waveform conforming to the designed covariance is synthesized while additionally enforcing inter-user interference suppression via the zero-forcing approach and imposing practical implementation constraints, notably limiting the \acrlong{par} on the individual antenna elements.
Simulation results demonstrate that the approach significantly enhances \gls{isac} waveform design flexibility, performance, and computational efficiency compared to the alternative methods in the literature.
\end{abstract}
\color{black}

\glsresetall

%
\IEEEpeerreviewmaketitle

\section{Introduction}
\label{Sec:Intro}
\IEEEPARstart{I}n \gls{isac} systems, the distinct requirements of sensing and communication are unified within a single joint waveform design framework, allowing for a controllable trade-off between sensing and communication goals \cite{liu2023seventy}. 
These requirements generally include better target detection and estimation performance for sensing \cite{he2012waveform}, and reliable communication between users \cite{tse2005fundamentals}. The desirable waveform properties can be understood from the individual treatment of radar and communication systems.

\gls{mimo} technology is one of the key elements in both modern wireless communication and radar systems \cite{tse2005fundamentals, he2012waveform}. 
In mobile \gls{mumimo} systems, a multi-antenna \gls{bs} typically serves multiple users simultaneously in the same frequency band.
Due to low computational power available at the end-user devices, the interference suppression should be implemented at the \gls{bs} \cite{wiesel2008zero}, 
making \gls{tx} precoding design in the downlink crucial for reliable communication \cite{joham2005linear}. 
Due to their lower complexity and good performance, linear \gls{zf} and \gls{mmse} precoders are typically employed in practice \cite{liu2023seventy}.

\add{
To enhance target detection and estimation performance, the main goal of \gls{mimo} radars is to transmit high signal power toward directions of interest, thereby ensuring strong illumination of anticipated targets and, consequently, improved target \gls{snr}.}
On the other hand, the cross-correlations between signals transmitted towards different targets should be low for better performance of radar \gls{rx} signal processing algorithms \cite{he2012waveform}. Moreover, a radar system should support null steering in interference directions to increase resilience against jammers and clutter \cite{van1988beamforming}. \looseness=-1

\add{
Waveform design for \gls{mimo} radar systems is performed using either a two-stage or a one-stage approach.
In the two-stage approach, a waveform covariance matrix producing a desired radiation pattern is constructed first, leveraging the fact that the \gls{tx} beampattern is fully determined by this matrix \cite{fuhrmann2008transmit}.
In the subsequent stage, a waveform conforming to the designed covariance matrix is synthesized \cite{stoica2008waveform}. In contrast, one-stage approaches aim to design the \gls{tx} waveforms directly, without the intermediate covariance matrix construction step \cite{ahmed2014mimo,li2017fast}. While one-stage designs offer a more direct formulation, two-stage approaches typically provide a greater flexibility to incorporating practical constraints on the waveform \cite{he2012waveform}.

Both one-stage and two-stage approaches measure the quality of the synthesized \gls{tx} beampattern using metrics such as \gls{isl}, \gls{psl}, \gls{ismr}, and magnitudes of the ripples within the illumination region. These metrics are commonly adopted as optimization objectives in various \gls{mimo} radar and \gls{isac} beamforming studies \cite{hua2013mimo, xu2015colocated, aubry2016mimo, fan2018constant, raei2022mimo, guo2024transmit}.
An alternative approach is to specify a desired \gls{tx} beampattern and minimize the deviation of the synthesized solution from this reference. This approach, commonly referred to as \gls{bpm} \cite{stoica2007probing}, does not directly optimize the classical quality metrics such as \gls{isl} or \gls{psl}. However, it offers better versatility by allowing the designer to freely define the desired \gls{tx} radiation pattern according to system-level requirements. Therefore, the \gls{bpm} criterion is extensively adopted in \gls{mimo} radar waveform design \cite{fuhrmann2008transmit, aittomaki2007signal, khabbazibasmenj2014efficient, zhang2015mimo, fan2018constant, zhang2022min} and in \gls{isac} beamforming applications \cite{liu2018mu, wu2022mimo, hua2023optimal}. \looseness=-1

Several studies further enhance the \gls{bpm} framework by incorporating cross-correlation minimization for improved radar \gls{rx} performance \cite{stoica2007probing, cheng2017constant, hammes2020generalized}. While originally developed for \gls{mimo} radar, the waveform covariance design methodology introduced in \cite{stoica2007probing} is widely adopted in \gls{isac} beamforming studies, with necessary adaptations to accommodate communication constraints, such as, user-specific minimum \gls{sinr} thresholds \cite{liu2020joint, zhang2022holographic}. The standard formulation typically results in a quadratic \gls{sdp}, with solution methods relying on general-purpose interior-point solvers. Consequently, the key drawback of these approaches lies in their high computational burden and limited scalability to high-dimensional problem instances \cite{malick2004dual}.
}

Another common approach is to design a reference sensing waveform yielding the desired \gls{tx} radiation characteristics and defining the distance to the reference sensing waveform, i.e., similarity metric \cite{liu2018toward, wang2024robust}, as the objective of minimization.
The typical communication performance objective introduced to this approach is the minimization of \gls{mui} energy \cite{liu2018toward,tsinos2021joint,guo2024transmit,wang2024robust,zhang2024transmit}, representing a special case of the \gls{mmse} precoder for \gls{mumimo} communication \cite{vtcpaper_bk}.
However, these approaches \cite{liu2018toward, wang2024robust} still require a waveform covariance input from \cite{fuhrmann2008transmit} to generate the reference sensing waveform for \gls{mimo} radar capabilities, which still relies on computationally intensive solvers.

Other sensing performance criteria have also been adopted for \gls{isac} waveform design, such as \gls{crb} for target estimation \cite{liu2021cramer,yang2023waveform} and radar \gls{sinr} \cite{tsinos2021joint, wen2023efficient}.
However, these approaches do not directly synthesize the desired beampattern, leading to a reduced flexibility in incorporating the expected target position uncertainty, which should be reflected in the beamwidth of the desired beampattern \cite{stoica2007probing}.

Finally, the \gls{tx} waveforms must also comply with practical constraints, namely, the available \gls{tx} power and the limitations of the \gls{rf} front-end.
To this end, it is desirable to enforce equal power distribution across all antennas by imposing a per-antenna power constraint \cite{yu2007transmitter}, and the \gls{tx} signal should be confined to the limited dynamic range supported by the \gls{pa} and \gls{dac}, to avoid non-linear distortion \cite{patton2008modulus}.
In addition to the operational constraints of the front-end, individual array elements can also fail \cite{agrawal1999active}, requiring the adaptation of the precoding algorithms to minimize the impact on the performance.
All of these factors make the adjustable \gls{par} and flexible control over individual antenna elements desirable for \gls{tx} waveform design.
Although some previous \gls{isac} studies incorporated the \gls{par} constraints \cite{bazzi2023integrated, wen2023efficient, yang2023waveform}, none of them can support control over individual antenna elements. 

\add{
This paper presents a comprehensive end-to-end waveform design framework for multi-antenna \gls{isac} systems, combining \gls{mumimo} communication and \gls{mimo} radar sensing. We adopt a two-stage design approach: the first stage focuses on the design of a \gls{tx} waveform covariance matrix for \gls{mimo} radar operations, followed by the synthesis of an \gls{isac} waveform in the second stage.

In the first stage, we address the waveform covariance matrix design for \gls{mimo} radar by introducing a novel convex matrix nearness formulation. This formulation aims to synthesize a desired spatial \gls{tx} power distribution while minimizing cross-correlations between \gls{tx} signals in different sensing directions.
It also enables interference suppression via null steering to mitigate jammer and clutter effects. Additionally, it allows for explicit lower and upper bounds on the \gls{tx} power of each antenna, thereby subsuming commonly used power constraints such as per-antenna and total power constraints as special cases. \looseness=-1

The resulting optimization problem is a quadratic \gls{sdp}. While general-purpose \gls{sdp} solvers can be employed to obtain a solution \cite{diamond2016cvxpy}, their computational complexity and poor scalability present limitations on practical feasibility \cite{malick2004dual}. To address this, we exploit the problem's inherent structure to develop a more efficient solution methodology. Specifically, we first establish a connection between the structure of the optimal covariance matrix and the dual variables of the problem. By exploiting the obtained result, we derive a computationally efficient first-order algorithm with the optimal convergence rate to solve the dual problem. The proposed algorithm also scales to large problem dimensions, rendering it suitable for massive \gls{mimo} scenarios. To the best of our knowledge, the proposed approach is the most computationally efficient covariance matrix design method supporting the aforementioned \gls{mimo} radar objectives. 

Finally, we propose a joint waveform design approach, in which the optimized covariance matrix serves as the foundation for \gls{mimo} radar operations. For the communication functionality, \gls{zf}-based \gls{mumimo} precoding is employed for the suppression of inter-user interference. The proposed joint waveform design framework facilitates an explicit trade-off between radar and communication performance, while simultaneously incorporating key practical constraints. Most notably, in contrast to existing methods, the framework permits independent control of the \gls{par} on each antenna, thereby providing greater flexibility for hardware implementation and system integration.
}

This work significantly extends and unifies our previous preliminary work on beamforming covariance matrix design \cite{globecom_bk} and waveform synthesis \cite{vtcpaper_bk}.
The proposed beamforming methodology is generalized to support arbitrary antenna array responses, flexible power allocation, and interference suppression through radiation null steering for improved robustness against jammer and clutter effects.
This unified approach not only consolidates the individual aspects addressed in our preliminary work but also introduces substantial improvements in performance and flexibility.

The remainder of this paper is organized as follows.
Section \ref{Sec:SystemModel} presents the system model, defines implementation constraints and describes the adopted communication and sensing design metrics. In Section~\ref{Sec:Proposed}, we present the overall proposed methodology. 
Numerical results are presented in Section \ref{Sec:Results} and the concluding remarks are given in Section \ref{Sec:Conclusions}.

The following notational convention is adopted in the paper.
\add{$\mathbb{C}^{M\times N}$ and $\mathbb{R}^{M\times N}$ denote the sets of $M \times N$ {complex and real matrices}, respectively.}
Matrices and vectors are denoted by upper- and lower-case bold letters, respectively.
$x_i$ denotes the $i$-th element of a vector $\boldsymbol{x}$, $\boldsymbol{x}_j$ denotes the $j$-th column of a matrix $\boldsymbol{X}$, $\boldsymbol{\Tilde{x}}_j^H$ denotes its $j$-th row, while $X_{ij}$ denotes the element in the $i$-th row and $j$-th column of $\boldsymbol{X}$.
$\Vert \boldsymbol{x} \Vert$ denotes the $\ell_2$-norm of $\boldsymbol{x}$, {and $\Vert \boldsymbol{x} \Vert_\infty = \max_i |x_i|$ denotes the $\ell_\infty$-norm of $\boldsymbol{x}$}.
$\Vert \boldsymbol{X} \Vert_F$ and $\Vert \boldsymbol{X} \Vert$ denote the Frobenius norm and the spectral norm of $\boldsymbol{X}$, respectively. Transpose, conjugate and conjugate-transpose (Hermitian) of $\boldsymbol{X}$ are denoted by $\boldsymbol{X}^T$, $\boldsymbol{X}^*$ and $\boldsymbol{X}^H$, respectively, while its trace is denoted by $\mathrm{Tr}(\boldsymbol{X})$. 
$\mathbb{H}^N$ is the set of Hermitian matrices of size $N$.
$\boldsymbol{X} \succeq \boldsymbol{0}$ denotes the positive semi-definiteness of a (Hermitian) matrix $\boldsymbol{X}$.  
$\boldsymbol{I}_N$ is the identity matrix of size $N$. $|c|$ is the modulus of $c\in \mathbb{C}$. 
$\mathrm{diag}\left(\{{x_i}\}_{i=1}^{N} \right)$ denotes an $N \times N$ diagonal matrix with $i$-th diagonal $x_i$. $\mathbb{E}[\,.\,]$ denotes statistical expectation. 

\section{System Model} 
\label{Sec:SystemModel}
In this section, we outline the signal model, describe the practical constraints on the design variables and present the adopted design goals for sensing and communication. 

\subsection{Signal Model}
\label{Sec:SignalModel}
\begin{figure}[t]
    \centering
    \includegraphics[width=0.75\columnwidth]{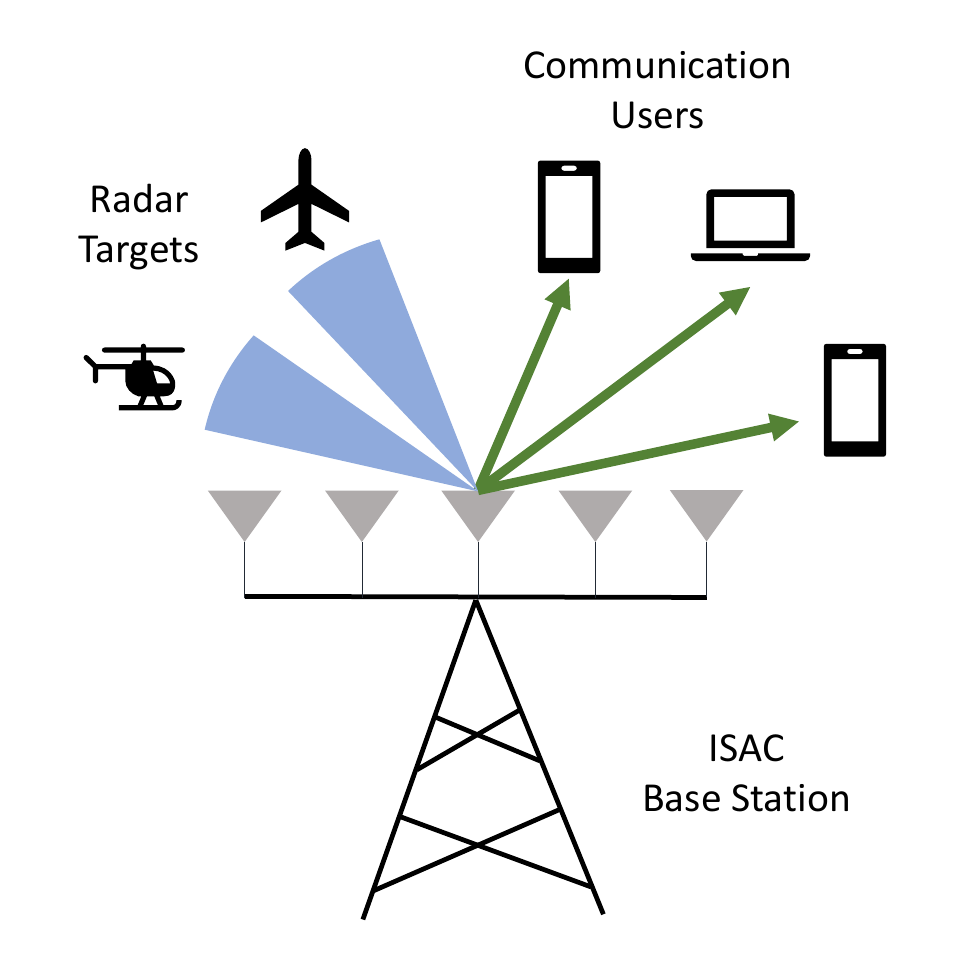}
    \caption{Illustration of the considered \gls{isac} system.}
    \label{system_model}
\end{figure}

\add{We consider a single-carrier, narrowband \gls{isac} system, where a \gls{bs} deploys a linear array with $M_T$ \gls{tx}-antennas. {Let $K_C$ denote the number of communication users}. As illustrated in Fig.~\ref{system_model}, \gls{bs} employs \gls{mimo} radar capabilities while simultaneously performing \gls{mumimo} downlink communication with $K_C < M_T$ single-antenna users. }

Let $\boldsymbol{x}[n] \in \mathbb{C}^{M_T \times 1}$ denote dual-functional waveform emitted by the \gls{bs} antenna array at time instance $n$ within the signaling block with length of $N_t$, $1 \le n \le N_t$, with $N_t > M_T$.  
\add{We note that $\boldsymbol{x}[n]$ denotes the \gls{tx} signal, which is to be designed to jointly support both communication and sensing functionalities.}
The corresponding complex baseband \gls{rx} signal at the $i$-th communication user can be written as \cite{tse2005fundamentals} 
\begin{equation}
    \label{Eqn:SingleUser}
    y_i[n] = \sqrt{\beta_i}\boldsymbol{\Tilde{h}}_i^H \boldsymbol{x}[n] + w_i[n], 
\end{equation}
where $\sqrt{\beta_i}\boldsymbol{\Tilde{h}}_i \in \mathbb{C}^{M_T \times 1}$ denotes the vector channel between the \gls{bs} and the user, where the large-scale fading effects are embedded in $\sqrt{\beta_i}$, and $w_i[n] \sim \mathcal{CN}({0}, \sigma_C^2)$ are i.i.d. additive white Gaussian noise samples.  
We also assume that the \gls{csi} is available at the \gls{tx}, e.g., obtained through feedback from the \gls{rx}. 

By jointly considering all $K_C$ users, we can write
\begin{equation}
    \label{Eqn:MultiUser}
    \boldsymbol{y}_C[n] = \boldsymbol{\Gamma H} \boldsymbol{x}[n] + \boldsymbol{w}_C[n], 
\end{equation}
where $\boldsymbol{\Gamma}=\mathrm{diag}\left(\{\sqrt{\beta_i}\}_{i=1}^{K_C}\right)$, $\boldsymbol{H}=[\boldsymbol{\Tilde{h}}_1~...~\boldsymbol{\Tilde{h}}_{K_C}]^H$, and $\boldsymbol{y}_C[n], \boldsymbol{w}_C[n]\in \mathbb{C}^{K_C \times 1}$ are obtained by stacking samples $\{y_i[n]\}_{i=1}^{K_C}$ and $\{w_i[n]\}_{i=1}^{K_C}$, respectively.
By considering a signal block length $N_t$ much shorter than the channel coherence time, the \gls{csi} is practically constant and \eqref{Eqn:MultiUser} can be extended as \looseness=-1
\begin{equation}
    \label{Eqn:MultiUserMultiSnapshot}
    \boldsymbol{Y}_C = \boldsymbol{\Gamma H} \boldsymbol{X} + \boldsymbol{W}_C, 
\end{equation}
where $\boldsymbol{Y}_C, \boldsymbol{W}_C \in \mathbb{C}^{K_C\times N_t}$ and $ \boldsymbol{X} \in \mathbb{C}^{M_T\times N_t}$ are obtained by a column-wise concatenation of vectors $\boldsymbol{y}_C[n]$, $\boldsymbol{w}_C[n]$, and $\boldsymbol{x}[n]$, respectively.

Different communication channel models can be adopted depending on the propagation scenario \cite{tse2005fundamentals}.
If the \gls{los} path between the \gls{bs} and the $i$-th user is obstructed, only scattered signal components arrive at the \gls{rx} and Rayleigh fading channel model is adopted, i.e.
\begin{equation}
    \label{Eqn:NlosChannel}
    \boldsymbol{\Tilde{h}}_i = \boldsymbol{\Tilde{h}}_{sc} \sim \mathcal{CN}({0}, \boldsymbol{I}_{M_T}).
\end{equation}
If the \gls{los} is unobstructed, Rician channel model is used, i.e.
\begin{equation}
    \label{Eqn:LosChannel}
    \boldsymbol{\Tilde{h}}_i = \sqrt{\frac{K}{K+1}} e^{j\psi_i}\boldsymbol{a}(\phi_i) + \sqrt{\frac{1}{K+1}} \boldsymbol{\Tilde{h}}_{sc},
\end{equation}
where $K$ is the Rician factor, $\psi_i \sim U[0,2\pi]$ is the random phase of the \gls{los} component, and $\phi_i\in [-\pi/2,\pi/2]$ denotes the azimuth angle, i.e., relative to the array boresight, at which the \gls{bs} observes the user, and $\boldsymbol{a}(\phi_i)$ is the corresponding array steering vector%
\footnote{In this work we assume that signals propagate in the horizontal plane, which is a reasonable assumption when the propagation distances are much larger than the antenna heights.
We further assume that scatterers in the communication channel (and radar targets) are in the far fields of the \gls{bs} antenna array.}.

With an appropriate calibration of the \gls{tx} array assumed, $\boldsymbol{a}(\phi_i)$ is known for a given $\phi_i$. 
For the widely considered special case of \glspl{ula} with half-wavelength inter-element spacing and no mutual coupling between the elements, it has the following form
\begin{equation}
    \label{Eqn:SteeringVector}
    \boldsymbol{a}(\phi_i)= [1~e^{j\pi \sin\phi_i}~...~e^{j(M_T-1) \pi \sin\phi_i }]^T.
\end{equation}

\add{
From the radar perspective, the \gls{tx} beampattern characterizes the spatial distribution of radiated power as a function of the direction $\phi$, defined as \cite{he2012waveform}
\begin{equation}
   \label{Eqn:TransmitBeampattern}
    P(\phi)
    = \boldsymbol{a}^H(\phi) \boldsymbol{R}\boldsymbol{a}(\phi), 
\end{equation}}where $\boldsymbol{R}=\mathbb{E}\left[\boldsymbol{x}[n]\boldsymbol{x}^H[n]\right]$ is the waveform covariance matrix, often represented by its sample covariance surrogate, i.e.,  
\begin{equation}
    \label{Eqn:CovMtxDefn}
    \boldsymbol{\hat{R}} =  \frac{1}{N_t}\boldsymbol{XX}^H.
\end{equation}

For colocated \gls{tx} and \gls{rx} with identical antenna array structures, the signal backscattered from $K_R$ point targets at angular directions $\phi_i,\phi_2,...,\phi_{K_R}$ located in the same range bin, 
can be written as \cite{he2012waveform}
\begin{equation}
    \label{Eqn:RadarRx}
    \boldsymbol{y}_R[n] = \sum_{i=1}^{K_R} q(\phi_i) \boldsymbol{a}^*(\phi_i)\boldsymbol{a}^H(\phi_i)\boldsymbol{x}[n] + \boldsymbol{w}_R[n],  
\end{equation}
where $q(\phi_i) \in \mathbb{C}$ includes propagation and scattering losses for $i$-th target, and $\boldsymbol{w}_R[n] \sim \mathcal{CN}({0}, \sigma_R^2\boldsymbol{I}_{M_T})$ denotes the i.i.d. additive complex Gaussian noise samples. 
Assuming that $q(\phi_i)$'s remain practically fixed within the signaling block $N_t$, \eqref{Eqn:RadarRx} can be extended to the whole signaling block as
\begin{equation}
    \label{Eqn:RadarRxMultiSnapshot}
    \boldsymbol{Y}_R = \sum_{i=1}^{K_R} q(\phi_i)\boldsymbol{a}^*(\phi_i)\boldsymbol{a}^H(\phi_i)\boldsymbol{X} + \boldsymbol{W}_R,  
\end{equation}
where $\boldsymbol{Y}_R, \boldsymbol{W}_R \in \mathbb{C}^{M_T\times N_t}$ are obtained by a column-wise concatenation of vectors $\boldsymbol{y}_R[n]$, $\boldsymbol{w}_R[n]$, respectively.

\subsection{Hardware Implementation Constraints} 
\label{Subsec:HardwareImp}
Apart from the goal to satisfy communication and sensing design criteria discussed in Sections \ref{Sec:SignalModelSensing} and \ref{Sec:SignalModelCommunication}, the signal waveform also has to meet some practical constraints discussed as follows. Let $\mathcal{R}$ and $\mathcal{X}$ denote the sets of $\boldsymbol{R}$ and $\boldsymbol{X}$ satisfying the given constraints, respectively.

With the reasonable goal to utilize all available energy, the most general constraint set for $\boldsymbol{R}$ would assign lower and upper \gls{tx} power bounds on each antenna element \cite{song2019fully}, i.e.,   
\begin{align}
    \mathcal{R}_{gen} = &\{ \boldsymbol{R}\in \mathbb{H}^{M_T} |~ \boldsymbol{R} \succeq \boldsymbol{0}, ~\mathrm{Tr}(\boldsymbol{R}) = P_T, 
    \nonumber \\   &l_i \le R_{ii} \le u_i, ~ 1\le i \le M_T\},
    \label{Eqn:R_gen}
\end{align}
where $u_i \ge l_i \ge 0$ and $\sum_{i=1}^{M_T} l_i \le P_T \le \sum_{i=1}^{M_T} u_i$, with $P_T>0$ denoting the total available \gls{tx} power. When the last constraint is removed from $\mathcal{R}_{gen}$, we obtain the total \gls{tx} power constraint set $\mathcal{R}_{tot}$. If all \gls{tx} antenna array elements are constrained to transmit with equal power $P_e=P_T/M_T$, we set $u_i=l_i=P_e$ in \eqref{Eqn:R_gen} to obtain the equal per-antenna power constraint set $\mathcal{R}_{pa}$. \looseness=-1 

Albeit stricter, the equal per-antenna power constraint is practically more desirable than the total power constraint \cite{fuhrmann2008transmit}. The per-antenna power constraint can be relaxed to allow for a pre-defined tolerable deviation factor $\delta > 0$ resulting in the constraint set $\mathcal{R}_{rel}$ \cite{aubry2016mimo}, obtained as the special case of $\mathcal{R}_{gen}$ by setting  $l_i = (1-\delta)P_e$ and $u_i = (1+\delta)P_e$ in \eqref{Eqn:R_gen}.

The temporal variation characteristic of the signal waveform is not considered by the previous sets, while being very important in practice due to the impacts on the efficiency and robustness of the amplifying stage in the \gls{tx}.
An important practical constraint is that on the \gls{par} \cite{he2012waveform}, defined as the ratio of the maximum \gls{tx} power to the average one, i.e.
\begin{equation}
    \label{Eqn:PAR}
    \mathrm{PAPR}(\boldsymbol{\Tilde{x}}_{i}) = \frac{\Vert \boldsymbol{\Tilde{x}}_{i} \Vert_{\infty}^2} { \Vert \boldsymbol{\Tilde{x}}_{i} \Vert^2/N_t },
\end{equation}
where $\boldsymbol{\Tilde{x}}_{i}^H$ denotes the $i$-th row of $\boldsymbol{X}$. By combining the constraints on \gls{par} and on individual antenna powers, imposed as
$\Vert \boldsymbol{\Tilde{x}}_{i} \Vert^2 = P_iN_t$ where $P_i$ is the power allocated for $i$-th antenna satisfying $\sum_i P_i = P_T$, the corresponding set is given as \looseness=-1
\begin{align}
    \label{Eqn:Xp}
    \mathcal{X}_{papr} = \{ \boldsymbol{X}=[\boldsymbol{\Tilde{x}}_1~...~\boldsymbol{\Tilde{x}}_{M_T}]^H
    \in \mathbb{C}^{M_T \times N_t} &|~ \nonumber \\ 
    \Vert \boldsymbol{\Tilde{x}}_{i} \Vert^2 = P_iN_t, \Vert \boldsymbol{\Tilde{x}}_{i} \Vert_{\infty}^2\le \rho_iP_i, ~ 1\le i & \le M_T\}.
\end{align}
We note that $1 \le \mathrm{PAPR}(\boldsymbol{\Tilde{x}}_{i}) \le N_t$, as $\Vert \boldsymbol{\Tilde{x}}_{i} \Vert_{\infty} \le \Vert \boldsymbol{\Tilde{x}}_{i} \Vert\le \sqrt{N_t}\Vert \boldsymbol{\Tilde{x}}_{i} \Vert_{\infty}$ for any $\boldsymbol{\Tilde{x}}_{i}$, and lower \gls{par} is desirable for higher power efficiency and/or lower signal distortion. 
This ensures $\mathrm{PAPR}(\boldsymbol{\Tilde{x}}_{i})\le \rho_i$, where $\rho_i\in [1,N_t]$ is the maximum \gls{par} supported by $i$-th antenna's \gls{rf} chain. In the limit when $\rho_i=1$, $\mathcal{X}_{papr}$ reduces to \gls{cm} constraint, i.e., $| {\Tilde{x}}_{i}[n] |^2 = P_i$ is constant for all $1 \le n \le N_t$.
\add{
On the other hand, by removing the $\ell_{\infty}$ constraint from $\mathcal{X}_{papr}$ \eqref{Eqn:PAR} and setting $P_i=P_e$ for all $1 \le i \le M_T$, we obtain the equal per-antenna power constraint set $\mathcal{X}_{pa}$, which is the waveform counterpart of $\mathcal{R}_{pa}$.
}
Since the temporal information is lost in $\boldsymbol{R}$, \gls{par} and \gls{cm} constraints are not relevant for $\mathcal{R}$.  

\subsection{Design Metrics for Sensing}
\label{Sec:SignalModelSensing}
We can observe from \eqref{Eqn:TransmitBeampattern} that the \gls{tx} beampattern is determined by $\boldsymbol{R}$.
The typical goal in \gls{mimo} radar sensing is to first design the waveform covariance matrix $\boldsymbol{R} \in \mathcal{R}$ matching the desired beampattern and then synthesizing the waveform $\boldsymbol{X}\in \mathcal{X}$. 
To formulate the covariance matrix design problem, we define the set of grid points $\Phi_g = \{ \bar{\phi}_i \}_{i=1}^{N_{\phi}}$ over the supported angular range $[-\pi/2,\pi/2]$ with $N_{\phi} > M_T$ denoting the grid size. The mainlobe region set, $\Phi_m$, is defined as the subset of $\Phi_g$ in which we expect sensing targets to be located. This prior information can be obtained in adaptive radar applications, e.g., using the previous estimates of target locations \cite{kilic2022adaptive}.
Hence, higher \gls{tx} power is desired to be directed towards $\bar{\phi}_i\in \Phi_m$. The sidelobe region is represented by $\Phi_s$, where $\Phi_m \cup \Phi_s = \Phi_g$ and $\Phi_m \cap \Phi_s = \varnothing$. 

\begin{figure}[t]
    \centering
    \includegraphics[width=7 cm]{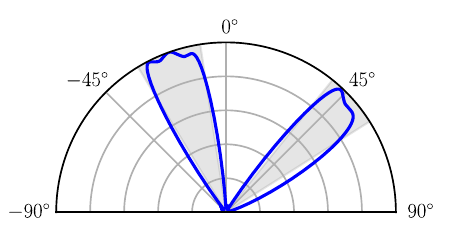}
    \caption{Illustration of the desired illumination directions (gray-shaded area) and generated beampattern (blue lines).}
    \label{polar_bp}
\end{figure}

Assume that every $\bar{\phi}_i \in \Phi_m$ is of the same importance,
then the (normalized) desired beampattern $P_d(\phi)$ is defined as
\begin{equation}
    \label{Eqn:DesiredBeampatternDefn}
    P_d(\bar{\phi}_i) =
    \begin{cases}
        1,~ &\mathrm{if}~ \bar{\phi}_i\in \Phi_m, \\
        0,~& \mathrm{otherwise},
    \end{cases}
\end{equation}
i.e., ideally we want to direct \gls{tx} signal exclusively towards mainlobe regions. 
Since the goal is to design $\boldsymbol{R}$ such that the generated $P(\phi)$ is close to $P_d(\phi)$, the following cost function can be adopted for \gls{bpm}
\begin{align}
    \label{Eqn:Jsb_}
    J^s_{\mathrm{bpm}}(\boldsymbol{R}, \eta) = \sum_{i=1}^{N_{\phi}} \left( \boldsymbol{a}^H(\bar{\phi}_i)\boldsymbol{R}\boldsymbol{a}(\bar{\phi}_i) - \eta P_d(\Bar{\phi}_i) \right)^2, 
\end{align}
where the scaling factor $\eta$ is introduced as $P(\phi)$ and $P_d(\phi)$ are scaled differently.
Fig. \ref{polar_bp} shows an example of generated and desired beampatterns. 
 
In many \gls{isac} studies, \eqref{Eqn:Jsb_} is adopted as the sensing cost to be minimized jointly over $\boldsymbol{R}$ and $\eta$, e.g., see \cite{wu2022mimo, hua2023optimal}. 
As \eqref{Eqn:Jsb_} is defined based on the \gls{ls} metric, it is referred to as \gls{ls} approach throughout the manuscript. Another common approach is based on the \gls{wls} metric, where $P_d(\bar{\phi}_i)$ in \eqref{Eqn:Jsb_} is weighted with the corresponding $\cos \bar{\phi}_i$ and minimized over $\boldsymbol{R}$ \cite{fuhrmann2008transmit}, adopted, e.g., in \cite{liu2018toward, wang2024robust}. 
Both formulations are solved via standard-interior point methods, resulting in $O(M_T^6)$ per-iteration complexity \cite{liu2020joint, malick2004dual}.

Apart from \gls{bpm} performance quantified by the cost \eqref{Eqn:Jsb_}, as elaborated in Section \ref{Sec:Intro}, the cross-correlations between signals transmitted in different directions are desired to be small in magnitude.
Hence, the cross-correlation beampattern $\boldsymbol{a}^H(\bar{\phi}_i)\boldsymbol{R}\boldsymbol{a}(\bar{\phi}_j)$ should have low-magnitude values for $\bar{\phi}_i \ne \bar{\phi}_j$ \cite{stoica2007probing}, and the following cost function can be specified
\begin{equation}
    \label{Eqn:Jsc_}
    J^s_{\mathrm{cc}}(\boldsymbol{R}) = \sum_{i=1}^{N_\phi}\sum_{j=1, j\ne i}^{N_\phi} \left| \boldsymbol{a}^H(\bar{\phi}_i)\boldsymbol{R}\boldsymbol{a}(\bar{\phi}_j) \right|^2. 
\end{equation}
It is also typical to adopt the weighted average of $J^s_{\mathrm{bpm}}(\boldsymbol{R}, \eta)$ and $J^s_{\mathrm{cc}}(\boldsymbol{R})$ as the sensing cost in the \gls{isac} literature \cite{liu2020joint, zhang2022holographic, tang2024dual}, and the resulting problem is again solved using standard interior-point methods, typically via general-purpose solvers. 

Additionally, it is desirable to have deep notches in the \gls{tx} beampattern at directions where interfering sources, such as jammers and clutters, are located \cite{stoica2011optimization, li2017fast}. Let $\Phi_J \subseteq \Phi_g$ denote the subset of (known) interference directions, then $|\boldsymbol{a}^H(\bar{\phi}_i)\boldsymbol{R}\boldsymbol{a}(\bar{\phi}_i)|$ is desired to be very small when $\bar{\phi}_i \in \Phi_J$, motivating the following cost 
\begin{align}
    \label{Eqn:Js2_}
    J^s_{\mathrm{is}}(\boldsymbol{R}) 
    = \sum_{\Bar{\phi}_i \in \Phi_J} \boldsymbol{a}^H(\Bar{\phi}_i) \boldsymbol{R} \boldsymbol{a}(\Bar{\phi}_i), 
\end{align}
which is defined to be linear in $\boldsymbol{R}$, unlike $J^s_{\mathrm{cc}}(\boldsymbol{R})$ \eqref{Eqn:Jsc_}, since $\boldsymbol{a}^H(\Bar{\phi}_i) \boldsymbol{R} \boldsymbol{a}(\Bar{\phi}_i) \ge 0$ by the positive semidefiniteness of $\boldsymbol{R}$. 

For the waveform design optimization, an appropriate cost function evaluates how closely the waveform $\boldsymbol{X}$ matches the designed $\boldsymbol{R}_0$. One obvious choice is to use $\Vert \boldsymbol{X}\boldsymbol{X}^H/N_t - \boldsymbol{R}_0 \Vert_F^2$, which is quartic in $\boldsymbol{X}$ and hence difficult to work with in general \cite{wang2024robust}. However, the class of $\boldsymbol{X}$'s satisfying $\boldsymbol{X}\boldsymbol{X}^H/{N_t}=\boldsymbol{R}_0$ can be written as 
\begin{equation}
     \label{Eqn:WaveformIdealSensing}
    \boldsymbol{X} = \sqrt{N_t} \boldsymbol{R}_0^{1/2}\boldsymbol{U},
\end{equation}
where $\boldsymbol{U} \in \mathbb{C}^{M_T \times N_t}$ is a semi-unitary matrix, i.e., $\boldsymbol{UU}^H = \boldsymbol{I}_{M_T}$. Hence, the following cost is proposed in \cite{stoica2008waveform}
\begin{equation}
    \label{Eqn:WaveformCostSensing}
    J_{\mathrm{syn}}^s(\boldsymbol{X}, \boldsymbol{U}) = \Vert \boldsymbol{X} - \sqrt{N_t} \boldsymbol{R}_0^{1/2}\boldsymbol{U} \Vert_F^2,
\end{equation}
where $\boldsymbol{U}$ is defined as an additional optimization variable.
We can interpret $\boldsymbol{U}$ as the semi-unitary factor adapting to the constraints imposed on $\boldsymbol{X}$ as $\boldsymbol{X}\in\mathcal{X}$ must be incorporated in the sequel. 

A simpler cost on sensing waveform can be defined based on the similarity metric, i.e.,  
\begin{equation}
    \label{Eqn:SimilarityMetric}
    J_{\mathrm{ref}}^s(\boldsymbol{X}) = \Vert \boldsymbol{X}-\boldsymbol{X}_0 \Vert_F^2,
\end{equation}
where $\boldsymbol{X}_0$ is a selected reference sensing waveform \cite{tsinos2021joint, bazzi2023integrated}. 
$\boldsymbol{X}_0$ can also be selected using \gls{bpm} criterion \cite{liu2018toward, wang2024robust}, e.g., as in \eqref{Eqn:WaveformIdealSensing}, however, adopting $J_{\mathrm{syn}}^s(\boldsymbol{X}, \boldsymbol{U})$ \eqref{Eqn:WaveformCostSensing} fits better for the synthesis of $\boldsymbol{R}_0$, due to the introduction of the additional optimization variable $\boldsymbol{U}$ adapting to $\boldsymbol{X}\in\mathcal{X}$ \cite{vtcpaper_bk}.

\subsection{Design Metrics for Communication}
\label{Sec:SignalModelCommunication}
Let $\boldsymbol{s}[n] \in \mathbb{C}^{K_C \times 1}$ denote the symbol vector which we want to transmit to the users at time instance $n$, with symbols drawn from an arbitrary constellation, e.g., M-QAM. Each user estimates the \gls{tx} symbol from the \gls{rx} signal as 
\begin{equation}
    \hat{s}_i =  g y_i = {s}_i + \left(g\sqrt{\beta_i}\boldsymbol{\Tilde{h}}_i^H\boldsymbol{x}-{s}_i\right) + g w_i, 
\end{equation}
with $g \in \mathbb{R}$ denoting the \gls{rx} gain, assumed to be the same for all users here \cite{jacobsson2016nonlinear}. 
The goal of the precoder is to design the \gls{tx} signal vector  $\boldsymbol{x}$ and scalar $g$, such that the estimated symbols are close to the desired communication symbols, i.e., $g \boldsymbol{y} \approx \boldsymbol{s}$. \looseness=-1

Assuming that $\boldsymbol{w}_C$ is uncorrelated with $\boldsymbol{s}$ and $\boldsymbol{x}$, also assigning equal weights to each user \cite{khorsandmanesh2023optimized}, the sum \gls{mse} for all users can be written as
\begin{equation}
    \label{Eqn:WFcost_expected}
    \mathbb{E}\left[\Vert {g} \boldsymbol{y} - \boldsymbol{s}\Vert^2\right] 
    = \mathbb{E}[\Vert  {g}  \boldsymbol{\Gamma Hx} - \boldsymbol{s}\Vert^2] + g^2\sigma_C^2 K_C ,
\end{equation}
which defines the cost for the well-known \gls{wf} approach \cite{joham2005linear}. 
If $g$ is restricted to be the same within the whole block length, by replacing $\mathbb{E}\left[\Vert {g} \boldsymbol{\Gamma Hx} - \boldsymbol{s}\Vert^2\right]$ with its sample-based average, the following objective can be adopted for communication 
\begin{equation}
    \label{Eqn:WFcost}
    J^c_{\mathrm{wf}}(\boldsymbol{X}, {g}) = \Vert {g}\boldsymbol{ \Gamma HX}-\boldsymbol{S}\Vert_F^2 + {g} ^2\sigma_C^2 K_C N_t,
\end{equation}
where $\boldsymbol{S} \in \mathbb{C}^{K_C \times N_t}$ is obtained by the column-wise concatenation of vectors $\boldsymbol{s}[n]$, $1\le n \le N_t$. The introduction of $g$ requires the users to estimate that parameter when the constellation is not constant-modulus (e.g., 16-QAM) \cite{jacobsson2016nonlinear}\footnote{When the constellation is constant-modulus such as \gls{qpsk}, $g$ should still be optimized by the \gls{tx}, however, the users do not require that parameter for decoding.}. Since $g$ is restricted to be constant during the whole signaling block in the given formulation, pilot signals can be used to transmit $g$ to the users at the cost of a communication overhead, or $g$ can be estimated by the users blindly \cite{jacobsson2016nonlinear}. 

Some studies adopt \eqref{Eqn:WFcost} as the communication cost in the \gls{isac} literature \cite{liu2021constant, cheng2021transmit}, by defining $g$ as an additional optimization variable. However, most studies tacitly assume $g=1$ \cite{wen2023efficient} and hence adopt $\Vert \boldsymbol{\Gamma H X}-\boldsymbol{S}\Vert_F^2$, often referred to as the \gls{mui} cost, e.g., see \cite{liu2018toward,tsinos2021joint,guo2024transmit,wang2024robust,zhang2024transmit}. Note that by fixing $g=1$,  $\Vert \boldsymbol{\Gamma H X}-\boldsymbol{S}\Vert_F^2$ can be very large even though $ \boldsymbol{\Gamma H X}$ is perfectly proportional to $\boldsymbol{S}$ \cite{li2024waveform}, resulting in significant performance losses \cite{wen2023efficient, vtcpaper_bk}. 

In \gls{zf} approach, channel inversion is performed at \gls{tx} to null the interference, which is obtained by setting $J^c_{\mathrm{wf}}(\boldsymbol{X}, {g})=0$ for $\sigma_C=0$. In the unconstrained case, this results in 
\begin{equation}
    \label{Eqn:Zf_expression}
    \boldsymbol{X} = g^{-1} \boldsymbol{H}^{\dagger} \boldsymbol{\Gamma}^{-1}\boldsymbol{S},
\end{equation}
where $\boldsymbol{H}^{\dagger} = \boldsymbol{H}^H(\boldsymbol{H}\boldsymbol{H}^H)^{-1}$. Therefore, the following cost can be adopted 
\begin{equation}
    \label{Eqn:ZFcost}
    J^c_{\mathrm{zf}}(\boldsymbol{X}) = \Vert \boldsymbol{X}- g^{-1} \boldsymbol{H}^{\dagger}\boldsymbol{\Gamma}^{-1}\boldsymbol{S}\Vert_F^2.   
\end{equation}
When we impose $\boldsymbol{X}\in\mathcal{X}_{tot}$, we achieve $J^c_{\mathrm{zf}}(\boldsymbol{X}) = 0$ when
\begin{equation}
\label{Eqn:ZFtotPowerSoln}
\boldsymbol{X}=\frac{\boldsymbol{H}^{\dagger}\boldsymbol{\Gamma}^{-1}\boldsymbol{S}}{\Vert \boldsymbol{H}^{\dagger}\boldsymbol{\Gamma}^{-1}\boldsymbol{S} \Vert_F}\sqrt{N_tP_T},~g=\frac{\Vert \boldsymbol{H}^{\dagger}\boldsymbol{\Gamma}^{-1}\boldsymbol{S} \Vert_F}{\sqrt{N_tP_T}}.
\end{equation}
Note that if $g=1$ were assumed a priori, it would not be possible to achieve $J^c_{\mathrm{zf}}(\boldsymbol{X})=0$ while satisfying $\boldsymbol{X}\in \mathcal{X}_{tot}$. 

In general, \gls{wf} approach outperforms \gls{zf} given that $g$ in \eqref{Eqn:WFcost} is optimized properly \cite{joham2005linear}. On the other hand, as elaborated in Section \ref{Sec:JointWaveformDesign}, \eqref{Eqn:ZFcost} enables individual control over each antenna unlike \eqref{Eqn:WFcost} due to the decoupled structure of $\boldsymbol{\Gamma H}$ and $\boldsymbol{X}$, also making it possible to incorporate adjustable \gls{par} constraints.  

\section{Proposed Methodology}
\label{Sec:Proposed}
\subsection{Overview}

\begin{figure}[t]
    \centering
    \includegraphics[width=8.5 cm]{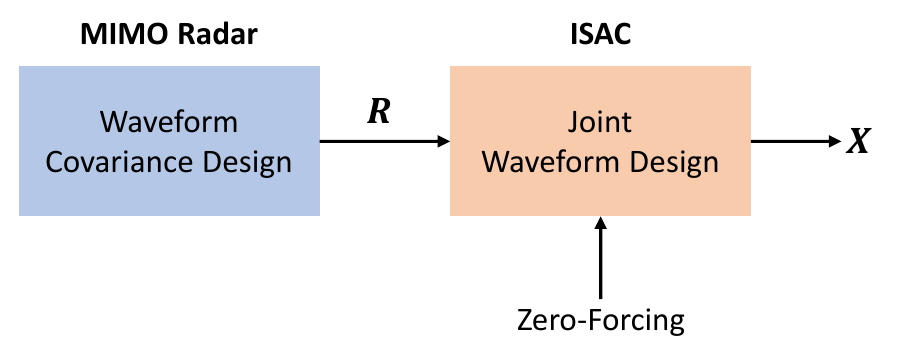}
    \caption{\add{Simplified flowchart of the proposed methodology.}}
    \label{diagram}
\end{figure}

\add{
In this section, we present the overall proposed methodology. As outlined in Section~\ref{Sec:Intro}, we first design a waveform covariance matrix tailored for \gls{mimo} radar operations. This matrix then serves as the input to a subsequent joint waveform design stage, which incorporates both radar and communication performance metrics.  

In Section~\ref{Sec:BeampatternSynthesis}, we present our waveform covariance design procedure, which targets three key objectives: beampattern synthesis, cross-correlation minimization, and null steering for interference suppression. Accordingly, we formulate the overall cost function by combining the corresponding objective terms: the \gls{bpm} cost $J^s_{\mathrm{bpm}}(\boldsymbol{R}, \eta)$ \eqref{Eqn:Jsb_}, the cross-correlation penalty $J^s_{\mathrm{cc}}(\boldsymbol{R})$ \eqref{Eqn:Jsc_}, and the interference suppression term $J^s_{\mathrm{is}}(\boldsymbol{R})$ \eqref{Eqn:Js2_}. We impose the constraint $\boldsymbol{R} \in \mathcal{R}_{gen}$, where $\mathcal{R}_{gen}$ is defined in \eqref{Eqn:R_gen}. This formulation allows for flexible control over the \gls{tx} power of each antenna by specifying individual lower and upper bounds. As discussed in Section~\ref{Subsec:HardwareImp}, the commonly used total power and equal per-antenna power constraints arise as special cases of this generalized hardware implementation constraint set.

In Section~\ref{Sec:JointWaveformDesign}, we introduce our joint waveform design strategy, which addresses two primary objectives: synthesizing the covariance matrix designed for sensing in the prior step and mitigating inter-user interference through \gls{zf}-based precoding. Accordingly, we define the overall cost function by combining the waveform synthesis cost $J_{\mathrm{syn}}^s(\boldsymbol{X}, \boldsymbol{U})$ \eqref{Eqn:WaveformCostSensing} and the \gls{zf} cost $J^c_{\mathrm{zf}}(\boldsymbol{X})$ \eqref{Eqn:ZFcost}. We impose the constraint $\boldsymbol{X} \in \mathcal{X}_{papr}$, as defined in \eqref{Eqn:PAR}. This formulation allows control over the \gls{par} on each antenna. As discussed in Section~\ref{Subsec:HardwareImp}, commonly used \gls{cm} and equal per-antenna power constraints are special cases of this hardware implementation constraint set.
The overall procedure is illustrated with a simple diagram in Fig. \ref{diagram}. 
}

\subsection{Waveform Covariance Design}
\label{Sec:BeampatternSynthesis}
Based on the beamspace representation of \gls{mimo} channels \cite{heath2016overview}, we define the matrix $\boldsymbol{D}$ as by concatenating the steering vectors corresponding to the selected angular grid points column-wise, i.e.,
\begin{align}
    \label{Eq:D}
    \boldsymbol{D} =\left[\boldsymbol{a}(\bar{\phi}_1)~ \boldsymbol{a}(\bar{\phi}_2)~...~\boldsymbol{a}(\bar{\phi}_{N_{\phi}})\right]\in \mathbb{C}^{M_T \times N_{\phi}},
\end{align}
assumed to be full-rank, which holds unless multiple grid points are assigned to the same angular direction. Beyond this, we impose no additional assumptions on $\boldsymbol{D}$. Consequently, the columns of $\boldsymbol{D}$ can represent the steering vectors of an arbitrary array with any spatial sampling (grid selection) configuration.  

We also define the diagonal matrix $\boldsymbol{T} \in \mathbb{R}^{N_\phi \times N_\phi}$ 
\begin{equation}
\label{Eqn:Tdefn}
\boldsymbol{T} = \eta \times \mathrm{diag}\left(\{P_d(\bar{\phi}_i)\}_{i=1}^{N_{\phi}} \right), 
\end{equation}
whose diagonals define the ideal \gls{tx} radiation pattern, while the off-diagonals correspond to the ideal zero cross-correlation case, allowing us to write 
\begin{equation}
    \label{Eqn:Js1_}
    J^s_\mathrm{bpm}(\boldsymbol{R},\eta) + J^s_\mathrm{cc}(\boldsymbol{R}) = \Vert \boldsymbol{D}^H\boldsymbol{R}\boldsymbol{D} - \boldsymbol{T} \Vert_F^2, 
\end{equation}
which defines the joint cost function for \gls{bpm} and the minimization of cross-correlations. As discussed in our preliminary work \cite{globecom_bk}, $\eta$ in \eqref{Eqn:Tdefn} can be used to control the importance of low cross-correlations between signals transmitted in different directions.
By also incorporating \eqref{Eqn:Js2_}, we define the overall cost as
\begin{equation}
    \label{Eqn:OveralCost1}
    J^s(\boldsymbol{R}) = \frac{1}{2}\Vert \boldsymbol{D}^H\boldsymbol{R}\boldsymbol{D} - \boldsymbol{T} \Vert_F^2 + \omega J^s_{\mathrm{is}}(\boldsymbol{R}), 
\end{equation}
where $\omega \ge 0$ is the weight assigned to interference suppression that should be selected based on the power of the interfering source, i.e., deeper notches are formed as $\omega$ gets larger. 

Considering the general constraint set $\mathcal{R}_{gen}$ \eqref{Eqn:R_gen}, we formulate covariance design as the following optimization problem 
\begin{align}
    \label{Eqn:ProposedOptCovMtx2}    \min_{\boldsymbol{R}}~& J^s(\boldsymbol{R}) \nonumber \\
    \mathrm{s.t.~} & \mathrm{Tr}(\boldsymbol{E}_i\boldsymbol{R}) \le u_i, ~ 1\le i \le M_T, \nonumber \\ 
    & -\mathrm{Tr}(\boldsymbol{E}_i\boldsymbol{R}) \le -l_i, ~ 1\le i \le M_T, \nonumber \\ 
    &\mathrm{Tr}(\boldsymbol{R})=P_T,~\boldsymbol{R} \succeq \boldsymbol{0}. 
\end{align}
In \eqref{Eqn:ProposedOptCovMtx2}, $\boldsymbol{R}\in \mathcal{R}_{gen}$ is expressed explicitly by defining the diagonal matrix $\boldsymbol{E}_i$ whose $i$-th diagonal is equal to one and zero elsewhere. By writing the \gls{svd} of $\boldsymbol{D}$ as $\boldsymbol{D}=\boldsymbol{U}_D\boldsymbol{\Sigma}_D\boldsymbol{V}_D^H$, where $\boldsymbol{U}_D\boldsymbol{U}_D^H=\boldsymbol{U}_D^H\boldsymbol{U}_D=\boldsymbol{I}_{M_T}$, $\boldsymbol{V}_D^H\boldsymbol{V}_D=\boldsymbol{I}_{M_T}$ and $\boldsymbol{\Sigma}_D\in \mathbb{R}^{M_T\times M_T}$, we define
\begin{equation}
    \label{Eqn:RtildeDefn}
    \boldsymbol{\tilde{R}} = (\boldsymbol{U}_D\boldsymbol{\Sigma}_D)^H\boldsymbol{R}(\boldsymbol{U}_D\boldsymbol{\Sigma}_D), 
\end{equation}
due to the full-rank assumption on $\boldsymbol{D}$, we can express $\boldsymbol{R}$ as \looseness=-1
\begin{equation}
    \label{Eqn:RandRtilde}
    \boldsymbol{R} = \boldsymbol{U}_{D} \boldsymbol{\Sigma}_{D}^{-1} \boldsymbol{\tilde{R}} \boldsymbol{\Sigma}_{D}^{-1} \boldsymbol{U}_{D}^H.  
\end{equation}
As shown in Appendix \ref{Sec:AppendixProgramEquality}, an equivalent problem to \eqref{Eqn:ProposedOptCovMtx2} is
\begin{align}
    \label{Eqn:ProposedOptCovMtx_simplified}
    \min_{\boldsymbol{\Tilde{R}}}~& \left\{ \frac{1}{2}\Vert \boldsymbol{\Tilde{R}} - \boldsymbol{\Tilde{T}} \Vert_F^2 +  \mathrm{Tr}\left(\boldsymbol{\Psi \Tilde{R}} \right) \right\} \nonumber \\
    \mathrm{s.t.~} & \mathrm{Tr}(\boldsymbol{\Tilde{E}}_i\boldsymbol{\Tilde{R}}) \le u_i, ~ 1\le i \le M_T, \nonumber \\ 
    & -\mathrm{Tr}(\boldsymbol{\Tilde{E}}_i\boldsymbol{\Tilde{R}}) \le -l_i, ~ 1\le i \le M_T, \nonumber \\ 
    &\mathrm{Tr}(\boldsymbol{C\Tilde{R}})=P_T,~\boldsymbol{\Tilde{R}} \succeq \boldsymbol{0}, 
\end{align}
where 
\begin{align}
    \label{Eqn:ProposedOptCovMtx_simplified_defns}
    &\boldsymbol{\Tilde{T}} = \boldsymbol{V}_D^H\boldsymbol{T}\boldsymbol{V}_D, \\
    &\boldsymbol{C} = \boldsymbol{\Sigma}_{D}^{-2},  \\
    &\boldsymbol{\Tilde{E}}_i = \boldsymbol{\Sigma}_{D}^{-1} \boldsymbol{U}_{D}^H \boldsymbol{E}_i  \boldsymbol{U}_{D} \boldsymbol{\Sigma}_{D}^{-1}, 
    \\
    &\boldsymbol{\Psi} = \omega\boldsymbol{\Sigma}_D^{-1} \boldsymbol{U}_D^H\left(\sum_{\Bar{\phi}_i \in \Phi_J} \boldsymbol{a}(\Bar{\phi}_i)\boldsymbol{a}^H(\Bar{\phi}_i) \right) \boldsymbol{U}_D \boldsymbol{\Sigma}_D^{-1}. 
\end{align}

The scaling of $\boldsymbol{\Tilde{R}}$ depends on $P_T$, while $\boldsymbol{\Tilde{T}}$ and $\boldsymbol{\Psi}$ may not. 
However, by setting $\eta$ in \eqref{Eqn:Tdefn} and $\omega$ in \eqref{Eqn:OveralCost1} proportional to $P_T$, the solution of \eqref{Eqn:ProposedOptCovMtx_simplified} becomes independent of $P_T$ (up to a scaling constant).

For the special case of $\boldsymbol{\Psi} = \boldsymbol{0}$, $\boldsymbol{\Tilde{E}}_i=\boldsymbol{E}_i$, $\boldsymbol{C}=\boldsymbol{I}_{M_T}$, and $u_i = l_i = P_e$, \eqref{Eqn:ProposedOptCovMtx_simplified} resembles the well-known nearest correlation matrix problem, widely encountered in finance applications \cite{higham2002computing, malick2004dual, boyd2005least}. 
The problem \eqref{Eqn:ProposedOptCovMtx_simplified} has a strictly convex objective and hence admits a unique minimizer.
Moreover, Slater condition automatically holds for \eqref{Eqn:ProposedOptCovMtx2}, therefore, there is no duality gap and we can obtain the optimal solution with a dual approach. To proceed, we define the operator $[.]_+$ which projects its argument onto the positive semidefinite cone, i.e, $[\boldsymbol{M}]_+ = \sum_{i} \mathrm{max}(0,\lambda_i) \boldsymbol{q}_i \boldsymbol{q}_i^H$ where $\boldsymbol{M}=\sum_i \lambda_i \boldsymbol{q}_i \boldsymbol{q}_i^H$ is obtained via the eigendecomposition of $\boldsymbol{M}$.

As derived in Appendix \ref{Sec:AppendixDualDerivation}, the dual of \eqref{Eqn:ProposedOptCovMtx_simplified} can be written as \looseness=-1
\begin{align}
    \label{Eqn:SimplifiedDualProb}
     \min_{\boldsymbol{\nu}\ge \boldsymbol{0}, \lambda} f(\boldsymbol{\nu}, \lambda)
\end{align}
where  
\begin{align}
    \label{Eqn:DualObj}
      f(\boldsymbol{\nu}, \lambda) = 
      \frac{1}{2} \big\Vert \boldsymbol{\tilde{R}}(\boldsymbol{\nu}, \lambda) \big\Vert_F^2 - \frac{1}{2}\Vert  \boldsymbol{\Tilde{T}} \Vert_F^2 & +\boldsymbol{\nu}^T\boldsymbol{b} + \lambda P_T, 
\end{align}
and
\begin{align}
\label{Eqn:CovMtxExpression}
& \boldsymbol{\tilde{R}}(\boldsymbol{\nu}, \lambda) = \big[\boldsymbol{\Tilde{T}} - \boldsymbol{E}(\boldsymbol{\nu}) - \boldsymbol{C}(\lambda)- \boldsymbol{\Psi} \big]_+, \\ 
\label{Eqn:DefinitionsDual1}
&\boldsymbol{\Bar{E}}_i =
\begin{cases}
    \boldsymbol{\Tilde{E}}_i,~ & 1\le i\le M_T, \\
    -\boldsymbol{\Tilde{E}}_i,~&M_T+1\le i\le 2M_T,
\end{cases}  \\
\label{Eqn:DefinitionsDual2}
&\boldsymbol{E}(\boldsymbol{\nu}) = \sum_{i=1}^{2M_T} \nu_i \boldsymbol{\Bar{E}}_i,~ \boldsymbol{C}(\lambda) = \lambda \boldsymbol{C},  \\ 
\label{Eqn:DefinitionsDual3}
&b_i =
\begin{cases}
    u_i,~ & 1\le i\le M_T, \\
    -l_i,~&M_T+1\le i\le 2M_T. 
\end{cases} 
\end{align}

The objective $f(\boldsymbol{\nu}, \lambda)$ in \eqref{Eqn:SimplifiedDualProb}  is (non-strongly) convex and also (once) differentiable with the following gradients \cite{boyd2005least}
\begin{align}
    \label{Eqn:PartialDerivatives}
    (\nabla_{\boldsymbol{\nu}}f)_i = b_i - \mathrm{Tr}\left( \boldsymbol{\tilde{R}}(\boldsymbol{\nu}, \lambda)\boldsymbol{\Bar{E}}_i \right), \nonumber \\ 
    \nabla_{\lambda}f = P_T- \mathrm{Tr}\left( \boldsymbol{\tilde{R}}(\boldsymbol{\nu}, \lambda)\boldsymbol{C} \right).   
\end{align}
Therefore, we can use a first-order method \cite{beck2017first} to solve \eqref{Eqn:SimplifiedDualProb}, which are appealing for their simplicity and low per-iteration complexity, yet constant step-size gradient schemes exhibit a convergence rate of $O(1/k)$ \cite{beck2017first}. Accelerated gradient (Nesterov's momentum) methods mitigate this limitation by achieving the optimal $O(1/k^2)$ rate, albeit without a monotonicity guarantee \cite{beck2009fast}.  
A widely adopted remedy is restarting the momentum upon detecting non-monotonic behavior, which enhances stability and performance across various applications \cite{o2015adaptive, giselsson2014monotonicity}. \looseness=-1

\begin{algorithm}[t]
	\caption{Procedure to solve \eqref{Eqn:SimplifiedDualProb}, based on accelerated projected gradient method with adaptive restart.}
	\begin{algorithmic}[1]
	\label{Algo:ProjectedGradient}
		\renewcommand{\algorithmicrequire}{\textbf{Input:}}
		\renewcommand{\algorithmicensure}{\textbf{Output:}}
        \REQUIRE $\boldsymbol{\Tilde{T}}$, $\boldsymbol{C}$, $\{ \boldsymbol{\Tilde{E}}_i \}_{i=1}^{M_T}$, $\boldsymbol{\Psi}$, $\boldsymbol{b}$, $P_T$
		\ENSURE $\boldsymbol{{\nu}^\star}$, ${{\lambda}^\star}$  
        \STATE {Initialize} $\boldsymbol{\nu}^0=\boldsymbol{{\nu}}^{-1}=\boldsymbol{0}$, $\lambda^0={\lambda}^{-1}=0$, $t^0= t^{-1} = 1$ 
        \STATE {For} $k\ge 0$ 
		\bindent
        \STATE $t^{k}=\big(1+\sqrt{1+4(t^{k-1})^2}\big)/2$
        \STATE $\boldsymbol{\Bar{\nu}}^{k} = \boldsymbol{{\nu}}^{k}+\big((t^{k-1}-1)/t^{k}\big)(\boldsymbol{{\nu}}^{k}-\boldsymbol{{\nu}}^{k-1})$
        \STATE ${\Bar{\lambda}}^{k} = {{\lambda}}^{k}+\big((t^{k-1}-1)/t^{k}\big)({{\lambda}}^{k}-{{\lambda}}^{k-1})$
        \STATE $\boldsymbol{{\nu}}^{k+1} = \left( \boldsymbol{\Bar{\nu}}^{k} - \gamma\nabla_{\boldsymbol{\nu}}f(\boldsymbol{\Bar{\nu}}^k, \Bar{\lambda}^k) \right)_+$
        \STATE  ${\lambda}^{k+1} = \Bar{\lambda}^{k}-\gamma\nabla_{\lambda} f(\boldsymbol{\Bar{\nu}}^k, \Bar{\lambda}^{k}) $
        \STATE {if} $(\boldsymbol{\Bar{\nu}}^k-\boldsymbol{\nu}^{k+1})^T(\boldsymbol{\nu}^{k+1}-\boldsymbol{\nu}^k)+({\Bar{\lambda}}^k-{\lambda}^{k+1})({\lambda}^{k+1}-{\lambda}^k) > 0$ 
        \bindentt
        \STATE $\boldsymbol{\Bar{\nu}}^{k}=\boldsymbol{{\nu}}^{k}$, ${\Bar{\lambda}}^{k}={{\lambda}}^{k}$
        \STATE $\boldsymbol{{\nu}}^{k+1} = \left( \boldsymbol{\Bar{\nu}}^{k} - \gamma\nabla_{\boldsymbol{\nu}}f(\boldsymbol{\Bar{\nu}}^k, \Bar{\lambda}^{k}) \right)_+$
        \STATE  ${\lambda}^{k+1} = \Bar{\lambda}^{k}-\gamma\nabla_{\lambda} f(\boldsymbol{\Bar{\nu}}^k, \Bar{\lambda}^{k})$
        \eindentt
        \eindent
	\end{algorithmic}
\end{algorithm}

To apply accelerated gradient updates with restart within our framework, we rewrite \eqref{Eqn:SimplifiedDualProb} in the standard composite form as
\begin{equation}
    \label{Eqn:SimplifiedDualIndicator}
    \min_{\boldsymbol{\nu}, \lambda}~ \{ f(\boldsymbol{\nu}, \lambda) + \mathcal{I}(\boldsymbol{\nu}) \},
\end{equation}
where $\mathcal{I}(\boldsymbol{\nu})$ is the indicator function of the nonnegative orthant 
\begin{equation}
    \label{Eqn:IndicatorFunction}
    \mathcal{I}(\boldsymbol{\nu}) = 
    \begin{cases}
        0~ & \mathrm{if} ~ \boldsymbol{\nu} \ge \boldsymbol{0}, \\
        \infty ~& \mathrm{otherwise}. 
    \end{cases}
\end{equation}
Furthermore, $f$ is $L$-smooth with \cite{boyd2005least}
\begin{equation}
    \label{Eqn:LipschitzConst}
    L=\sum_{i=1}^{2M_T} \Vert \boldsymbol{\Bar{E}}_i \Vert^2 + \Vert \boldsymbol{C} \Vert^2. 
\end{equation}

Let $\gamma$ denote the step size, then the proximal gradient update (without acceleration) for $\boldsymbol{\nu}$ and the gradient update for $\lambda$ at iteration number $k$ is  
\begin{align}
    \label{Eqn:nuUpdateProximal}
    \boldsymbol{\nu}^{k+1} = \mathrm{prox}_{\gamma\mathcal{I}}\left(\boldsymbol{\nu}^{k}-\gamma\nabla_{\boldsymbol{\nu}}f(\boldsymbol{\nu}^k, \lambda^k) \right),  \\
    {\lambda}^{k+1} = \lambda^{k}-\gamma\nabla_{\lambda} f(\boldsymbol{\nu}^k, \lambda^k),
\end{align}
where the $\mathrm{prox}$ operator is defined as \cite{beck2017first}
\begin{align}
     \label{Eqn:ProximalMapping}
     \mathrm{prox}_{\gamma\mathcal{I}}{(\boldsymbol{v})} = \argmin_{\boldsymbol{u}} ~\mathcal{I}(\boldsymbol{u}) + \frac{1}{2\gamma}\Vert \boldsymbol{u} - \boldsymbol{v} \Vert^2.
\end{align}
As $\mathcal{I}(.)$ is an indicator function defined in \eqref{Eqn:IndicatorFunction}, $\mathrm{prox}$ operator in \eqref{Eqn:ProximalMapping} reduces to projection onto the nonnegative orthant and $\boldsymbol{\nu}$ update in \eqref{Eqn:nuUpdateProximal} becomes the projected gradient update \cite{boyd2005least} 
\begin{align}
    \label{Eqn:nuUpdateProjection}
    \boldsymbol{\nu}^{k+1} = \left(\boldsymbol{\nu}^{k}-\gamma\nabla_{\boldsymbol{\nu}}f(\boldsymbol{\nu}^k, \lambda^k) \right)_+, 
\end{align}
where $(.)_+$ projects its argument onto the nonnegative orthant.


We summarize the solution procedure in Algorithm \ref{Algo:ProjectedGradient}, where Steps 3, 4, and 5 implement the momentum setup and the accelerated projected gradient updates are given in Steps 6 and 7. The restart scheme adopted in Step 8 of Algorithm \ref{Algo:ProjectedGradient} is based on the gradient mapping for $\boldsymbol{\nu}^{k+1}$ and ${\lambda}^{k+1}$ updates \cite{o2015adaptive}, i.e., Steps 6 and 7 of Algorithm \ref{Algo:ProjectedGradient}. 

Algorithm \ref{Algo:ProjectedGradient} is guaranteed to converge for $\gamma=1/L$ \cite{beck2017first}. When we restart the momentum, we do not restart $t^k$ as opposed to some studies on adaptive restart schemes, e.g., \cite{o2015adaptive}. This modification ensures  $O(1/k^2)$ convergence rate for our formulation \cite{giselsson2014monotonicity} since $f(\boldsymbol{\nu}, \lambda)$ in \eqref{Eqn:SimplifiedDualIndicator} is convex but not strongly convex. 

The stopping criterion for Algorithm \ref{Algo:ProjectedGradient} can be set as  
\begin{equation}
    \Vert \boldsymbol{\bar{\nu}}^{k} - \boldsymbol{\nu}^{k+1} \Vert/2M_T + |\Bar{\lambda}^{k}-\lambda^{k+1}|
    \le \varepsilon, 
\end{equation}
where $\varepsilon > 0$ is the stopping tolerance.  
After the optimal $\boldsymbol{\nu^\star}$ and $\lambda^\star$ are found by Algorithm \ref{Algo:ProjectedGradient}, the optimal $\boldsymbol{\Tilde{R}^\star} = \boldsymbol{\Tilde{R}}(\boldsymbol{\nu^\star}, \lambda^\star)$ is found by \eqref{Eqn:CovMtxExpression} and then the optimal desired waveform covariance matrix $\boldsymbol{R}^\star$ is computed using \eqref{Eqn:RandRtilde}.
\add{
Therefore, the optimal covariance matrix $\boldsymbol{R}^\star$ has the following structure
\begin{equation}
    \label{Eqn:RandRtilde_opt}
    \boldsymbol{R}^\star = \boldsymbol{U}_{D} \boldsymbol{\Sigma}_{D}^{-1}\big[\boldsymbol{\Tilde{T}} - \boldsymbol{E}(\boldsymbol{\nu}^\star) - \boldsymbol{C}(\lambda^\star)- \boldsymbol{\Psi} \big]_+ \boldsymbol{\Sigma}_{D}^{-1} \boldsymbol{U}_{D}^H.  
\end{equation}
}

\add{
Algorithm~\ref{Algo:ProjectedGradient} can be simplified under the total power ($\boldsymbol{R} \in \mathcal{R}_{{tot}}$) and equal per-antenna power ($\boldsymbol{R} \in \mathcal{R}_{{pa}}$) constraints. In these special cases, {\eqref{Eqn:ProposedOptCovMtx_simplified}} does not include any inequality constraints and the associated dual problem \eqref{Eqn:SimplifiedDualProb} consequently reduces to an unconstrained convex optimization problem. For instance, when the covariance matrix $\boldsymbol{R}$ is subject only to the total power constraint $\boldsymbol{R} \in \mathcal{R}_{{tot}}$, all computations involving $\boldsymbol{\nu}$ can be omitted from Algorithm \ref{Algo:ProjectedGradient}.
}

The computationally dominant operation in Algorithm \ref{Algo:ProjectedGradient} is the evaluation of the gradients in Steps 6 and 7, requiring the projection of $\big[ \boldsymbol{\Tilde{T}} - \boldsymbol{\Psi} - \boldsymbol{E}(\boldsymbol{\nu}) - \boldsymbol{C}(\lambda) \big]$ onto the positive semidefinite cone, which can be computed via the direct eigen-decomposition with complexity $O(M_T^3)$ \cite{higham2008functions}. 
Compared to the $O(M_T^6)$ per-iteration complexity of alternative approaches solved via standard interior-point methods, the proposed approach here provides significant computational gains.

\subsection{Joint Waveform Design}
\label{Sec:JointWaveformDesign}

The goal of joint waveform design is the unification of both communication and sensing criteria within a single cost.
This joint cost can be a weighted average of the sensing and communication costs described in Sections \ref{Sec:SignalModelSensing} and \ref{Sec:SignalModelCommunication}.
To synthesize the waveform, we formulate the following non-convex optimization problem combining the sensing cost \eqref{Eqn:WaveformCostSensing} and the communication cost \eqref{Eqn:ZFcost} as
\add{
\begin{align}
    \label{Eqn:ProposedJointDesign}
    &\min_{\boldsymbol{X},\boldsymbol{{U}}}~ J(\boldsymbol{X}, \boldsymbol{{U}}) \nonumber \\
    &\mathrm{s.t.~} \boldsymbol{X}\in \mathcal{X}_{papr},~ \boldsymbol{{U}} \boldsymbol{{U}}^H=\boldsymbol{I}_{M_T},  
\end{align}
where
\begin{equation}
    J(\boldsymbol{X}, \boldsymbol{{U}}) = \alpha J^c_{\mathrm{zf}}(\boldsymbol{X}) + (1-\alpha) J_{\mathrm{syn}}^s(\boldsymbol{X}, \boldsymbol{U}),
\end{equation}
}and $\alpha \in [0,1]$ controls the trade-off between sensing and communication performance%
\footnote{For $\alpha=1$ and $\alpha=0$, the cost function becomes exclusively communication- and sensing-oriented, respectively.}.

To solve \eqref{Eqn:ProposedJointDesign}, we adopt the \gls{am} algorithm by applying minimization over each variable by fixing the other \cite{beck2017first}.  

The minimization of $J(\boldsymbol{X}, \boldsymbol{{U}})$ over $\boldsymbol{{U}}$ is equivalent to solving the following non-convex problem
\begin{equation}
\label{Eqn:Procrustes}
\min_{\boldsymbol{{U}{U}}^H=\boldsymbol{I}_{M_T}}\Vert \boldsymbol{{X}} - \sqrt{N_t} \boldsymbol{{R}}_0^{1/2}\boldsymbol{{U}} \Vert_F^2,
\end{equation}
which is closely related with the orthogonal Procrustes problem \cite{higham2008functions}. We can equivalently write \eqref{Eqn:Procrustes} as 
\begin{equation}
\max_{\boldsymbol{{U}{U}}^H=\boldsymbol{I}_{M_T}}\mathrm{Re}\{ \mathrm{Tr}(\sqrt{N_t}\boldsymbol{{X}}^H\boldsymbol{{R}}_0^{1/2}\boldsymbol{{U}}) \},
\end{equation}
admitting the following closed-form solution 
\begin{equation}
    \label{Eqn:PolarDecomposition}
    \boldsymbol{{U}}^\star = \boldsymbol{\mathcal{U}}^H,
\end{equation}
where $\boldsymbol{\mathcal{U}}\in \mathbb{C}^{ N_t \times M_T}$ is the semi-unitary factor of the polar decomposition of $\sqrt{N}_t \boldsymbol{{X}}^H\boldsymbol{{R}}_0^{1/2}$ \cite[Ch. 8]{higham2008functions}.

The minimization of $J(\boldsymbol{X},\boldsymbol{{U}})$ over $\boldsymbol{X}$ is equivalent to solving 
\begin{align}
    \label{Eqn:WaveformDesignOriginal}
    \min_{{\boldsymbol{X}\in \mathcal{X}_{papr}}}~ \big\{ \alpha \Vert \boldsymbol{X}- g^{-1} \boldsymbol{H}^{\dagger}\boldsymbol{\Gamma}^{-1}\boldsymbol{S}\Vert_F^2 \nonumber \\ 
    + (1-\alpha) \Vert \boldsymbol{{X}} - &\sqrt{N_t} \boldsymbol{{R}}_0^{1/2}\boldsymbol{{U}} \Vert_F^2 \big\}.
\end{align}
For the simplicity of notation, we define $\boldsymbol{Z}=g^{-1} \boldsymbol{H}^{\dagger}\boldsymbol{\Gamma}^{-1}\boldsymbol{S}$ and $\boldsymbol{{G}}= \sqrt{N_t}\boldsymbol{{R}}_0^{1/2}\boldsymbol{{U}}$, which allows us to rewrite \eqref{Eqn:WaveformDesignOriginal} as
\add{
\begin{align}
    \label{Eqn:WaveformDesignSimplified}
    \min_{{\boldsymbol{X}\in \mathcal{X}_{papr}}}~ \big\{ \alpha \Vert \boldsymbol{X}- \boldsymbol{Z}\Vert_F^2  
    + (1-\alpha) \Vert \boldsymbol{{X}} - \boldsymbol{G} \Vert_F^2 \big\}.
\end{align}
}Due to proper normalizations of $\boldsymbol{G}$ and $\boldsymbol{Z}$ in \eqref{Eqn:WaveformDesignSimplified}, the value of $P_T$ only introduces a scaling factor to the solution. 
Furthermore $\beta_i$ in \eqref{Eqn:SingleUser} can be scaled by a common positive constant, without affecting the optimal solution.
This property allows generalization of the proposed method to diverse scenarios \cite{vtcpaper_bk}.

Let $\boldsymbol{\Tilde{g}}_i^H\in\mathbb{C}^{1 \times N_t}$ and $\boldsymbol{\Tilde{z}}_i^H\in\mathbb{C}^{1 \times N_t}$, $1 \le i \le M_T$, denote $i$-th row of $\boldsymbol{G}$ and $\boldsymbol{Z}$, respectively.
\add{
Given the row-separability of the objective function of \eqref{Eqn:WaveformDesignSimplified} and the constraint set $\mathcal{X}_{papr}$ \eqref{Eqn:Xp}, the optimization problem \eqref{Eqn:WaveformDesignSimplified} can be decoupled into $M_T$ independent subproblems, i.e.,
\begin{align}
    \label{Eqn:OptimizationWithPARConstraint_}
    \min_{\boldsymbol{\Tilde{x}}_i} ~ &\{ \alpha \Vert \boldsymbol{\Tilde{x}}_i - \boldsymbol{\Tilde{z}}_i \Vert^2 + (1-\alpha)\Vert \boldsymbol{\Tilde{x}}_i - \boldsymbol{\Tilde{g} }_i \Vert^2\} \nonumber \\ 
    \mathrm{s.t.~} &\Vert \boldsymbol{\Tilde{x}}_{i} \Vert^2 = P_iN_t, \Vert \boldsymbol{\Tilde{x}}_{i} \Vert_{\infty}^2\le \rho_iP_i. 
\end{align}
This decoupling enables parallel optimization such that $\boldsymbol{\Tilde{x}}_i$, representing the \gls{tx} signal by the $i$-th antenna, can be optimized independently. 
}
As $\Vert \boldsymbol{\Tilde{x}}_{i} \Vert^2 = P_iN_t$ is constant, \eqref{Eqn:OptimizationWithPARConstraint_} is equivalent to \cite{tropp2005designing}
\begin{align}
    \label{Eqn:OptimizationWithPARConstraint}
    \max_{\boldsymbol{\Tilde{x}}_i} \mathrm{Re}\{ \boldsymbol{\Tilde{x}}_i^H\boldsymbol{\gamma}_i  \} \mathrm{~~s.t.~~}  \Vert \boldsymbol{\Tilde{x}}_{i} \Vert^2 = P_iN_t,\;\Vert \boldsymbol{\Tilde{x}}_{i} \Vert_{\infty}^2\le \rho_iP_i,
\end{align}
where
\begin{equation}
    \boldsymbol{\gamma}_i = \alpha\boldsymbol{\Tilde{z}}_i + (1-\alpha)\boldsymbol{\Tilde{g}}_i.
\end{equation}

If the $\ell_{\infty}$-norm constraint is removed from \eqref{Eqn:OptimizationWithPARConstraint}, we obtain 
\begin{align}
    \label{Eqn:OptimizationWithPerConstraint}
    \max_{\boldsymbol{\Tilde{x}}_i} \mathrm{Re}\{ \boldsymbol{\Tilde{x}}_i^H\boldsymbol{\gamma}_i  \} \mathrm{~~s.t.~~}  \Vert \boldsymbol{\Tilde{x}}_{i} \Vert^2 = P_iN_t,
\end{align}
yielding the following closed-form solution  
\begin{equation}
    \label{Eqn:XsolnPerantenna}
    \boldsymbol{\Tilde{x}}^\star_i = \sqrt{{P_i N_t}}\frac{\boldsymbol{\gamma}_i}{\Vert \boldsymbol{\gamma}_i\Vert}, 
\end{equation}
which can be obtained via the Cauchy-Schwarz inequality.
When $\boldsymbol{X}\in\mathcal{X}_{pa}$, the solution for this special case is obtained by replacing $P_i = P_e$ in \eqref{Eqn:XsolnPerantenna} for all $i$. 

When the entries of $\boldsymbol{\Tilde{x}}_i$ are restricted to be of the same magnitude, which is the special case of \gls{par} constraint for $\rho_i=1$, i.e., $\boldsymbol{X} \in \mathcal{X}_{cm}$, \eqref{Eqn:OptimizationWithPARConstraint} becomes 
\begin{equation}
\label{Eqn:OptimizationWithCM}
    \max_{\boldsymbol{\Tilde{x}}_i} \mathrm{Re}\{ \boldsymbol{\Tilde{x}}_i^H\boldsymbol{\gamma}_i  \} \mathrm{~~s.t.~~}  | {\Tilde{x}}_{i}[n] |^2 = P_i,
\end{equation} 
where the closed-form solution can be obtained by aligning the phases of $\boldsymbol{\Tilde{x}}_i$ and $\boldsymbol{\gamma}_i$ with a proper scaling, i.e., 
\begin{equation}
    \label{Eqn:XsolnCM}
    \boldsymbol{\Tilde{x}}^\star_i = \sqrt{P_i} e^{  j\angle   \boldsymbol{\gamma}_i  },
\end{equation}
where $\angle$ returns the entry-wise phases of its argument.

While the special cases of \eqref{Eqn:OptimizationWithPARConstraint} yield closed-form solutions, the original problem itself does not have a closed-form solution. However, it can optimally be solved using the efficient recursive algorithm given in \cite[Alg.~2]{tropp2005designing}, which is widely adopted in radar applications \cite{he2012waveform}. 

The proposed joint waveform design method is summarized in Algorithm \ref{Algo:AlternatingMinimization}, where the stopping condition can be set as 
\begin{equation}
    \frac{\Vert \boldsymbol{X}^{k+1}-\boldsymbol{X}^{k}\Vert_F^2}{N_tP_T}\le \varepsilon. 
\end{equation}
The computationally dominant operation in this algorithm is the one in Step 2, involving the computation of the polar decomposition of $\sqrt{N}_t \boldsymbol{{X}}^H\boldsymbol{{R}}_0^{1/2}$, which is $O(M_T^2N_t)$ \cite{higham2008functions}. 

\begin{algorithm}[t]
	\caption{Procedure to solve \eqref{Eqn:ProposedJointDesign}, based on alternating minimization method.}
	\begin{algorithmic}[1]
	\label{Algo:AlternatingMinimization}
		\renewcommand{\algorithmicrequire}{\textbf{Input:}}
		\renewcommand{\algorithmicensure}{\textbf{Output:}}
		\REQUIRE $\boldsymbol{{R}}_0$, $\boldsymbol{H}$, $\boldsymbol{S}$, $\alpha$, $\mathcal{X}$
		\ENSURE $\boldsymbol{{X}^\star}$ \\
        \STATE Initialize $\boldsymbol{X}^0$ and set $g$ as given in \eqref{Eqn:ZFtotPowerSoln}; and for $k \ge 0$ \\
		\bindent
        \STATE Given $\boldsymbol{{X}}^k$, update $\boldsymbol{{U}}^{k+1}$ using \eqref{Eqn:PolarDecomposition}
        \STATE Given $\boldsymbol{{U}}^{k+1}$, update $\boldsymbol{\Tilde{x}}_i^{k+1}$ for all $i$
        \bindentt
        \STATE If $\mathcal{X}= \mathcal{X}_{papr}$, solve \eqref{Eqn:OptimizationWithPARConstraint} using \cite[Alg. 2]{tropp2005designing}
        \STATE If $\mathcal{X}= \mathcal{X}_{cm}$, use \eqref{Eqn:XsolnCM} 
        \STATE If $\mathcal{X}= \mathcal{X}_{pa}$, use \eqref{Eqn:XsolnPerantenna} for $P_i=P_e$
        \eindentt
        \STATE Set $\boldsymbol{X}^{k+1}=\left[\boldsymbol{\Tilde{x}}_1^{k+1}~...~\boldsymbol{\Tilde{x}}_{M_T}^{k+1}\right]^H$ 
        \eindent
	\end{algorithmic}
\end{algorithm}

\begin{figure*}%
    \centering
    \begin{subfigure}[t]{\figWidth cm}
        \includegraphics[width=\figWidth cm]{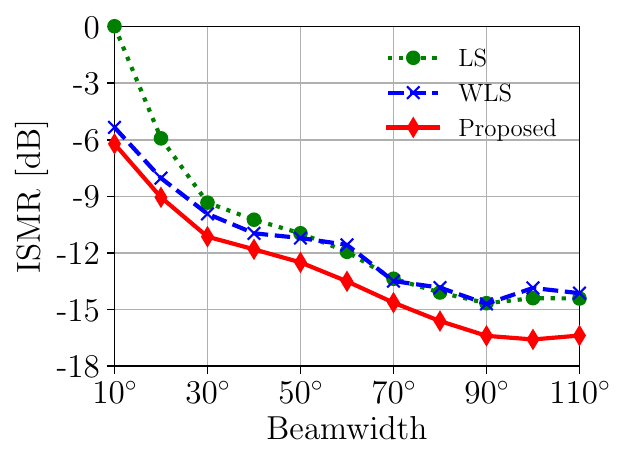}
        \subcaption{Integrated sidelobe to mainlobe ratios.}
        \label{ismr_changingbw}
    \end{subfigure}%
    \begin{subfigure}[t]{\figWidth cm}
        \includegraphics[width=\figWidth cm]{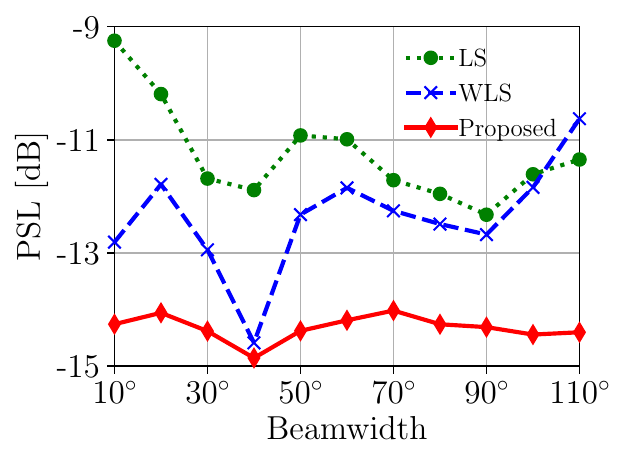}
        \subcaption{Peak sidelobe levels.}
        \label{psl_changingbw}
    \end{subfigure}%
    \begin{subfigure}[t]{\figWidth cm}
        \includegraphics[width=\figWidth cm]{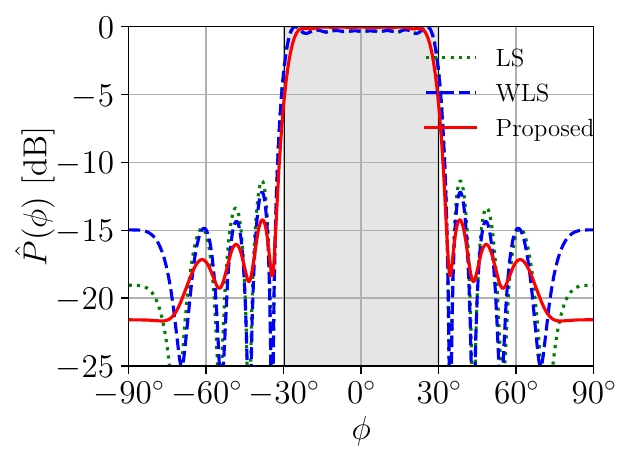}
        \subcaption{Synthesized beampattern ($60^{\circ}$ beamwidth).}
        \label{bp_changing_bw}
    \end{subfigure}%
    \caption{Synthesized beampattern quality as a function of desired beamwidth, when the mainlobe direction is $\phi = 0^{\circ}$.}
\end{figure*}

\section{Simulation Results}
\label{Sec:Results}
In this section we evaluate the performance of the proposed methods based on simulations,
We limit our analysis to the \gls{ula} to facilitate easier interpretation of results as it is the standard setup in the literature.
To this end, a \gls{ula} consisting of $M_T = 16$ antennas with half-wavelength spacing is considered.
A grid size of $N_{\phi}=2048$ is adopted (see Section~\ref{Sec:SignalModelSensing}), with grid points uniformly sampling the angular domain, i.e., the angle resolution is less than $0.1^{\circ}$. The total power budget\footnote{As discussed, the obtained waveform covariance matrix and waveform outputs of our proposed methods can simply be scaled for any $P_T$.}. is set to $P_T = 10$ W

\subsection{Performance Metrics}
For evaluation of the quality of solutions obtained by the proposed method, we adopt several standard metrics. For the proposed covariance matrix design method described in Section \ref{Sec:BeampatternSynthesis}, we adopt the widely-used \gls{psl} and \gls{ismr} metrics to measure the beampattern quality.
The level of power dissipation in unintended directions is typically measured by the \gls{psl}, evaluated relative to the maximum mainlobe power \cite{aubry2016mimo}.
The efficiency of synthesized patterns in focusing the available energy towards the desired directions is quantified by \gls{ismr}, computed as 
\begin{equation}
    \mathrm{ISMR} = \frac{\sum_{\Bar{\phi}_i \in \Phi_s} P(\Bar{\phi}_i)} {\sum_{\Bar{\phi}_j \in \Phi_m} P(\Bar{\phi}_j)}. 
\end{equation}

For visualization, we also present the normalized generated beampatterns, denoted as $\Hat{P}(\phi)$, for sample scenarios where each beampattern is normalized by the maximum beampattern value across all methods, ensuring that $0$ dB corresponds to the peak value of the generated beampatterns.

Apart from the beampattern synthesis performance, we also investigate the cross-correlation between signals transmitted in different mainlobe directions, defined as
\begin{equation}
    \label{Eq:Crosscorr}
    \rho_{i,j} = \frac{|\boldsymbol{a}^H(\bar{\phi}_i)\boldsymbol{R}\boldsymbol{a}(\bar{\phi}_j)|}{ \Vert \boldsymbol{R}^{1/2}\boldsymbol{a}(\bar{\phi}_i) \Vert \Vert \boldsymbol{R}^{1/2}\boldsymbol{a}(\bar{\phi}_j) \Vert  }, \quad \Bar{\phi}_i, \Bar{\phi}_j \in \Phi_m,
\end{equation}
and report the average value $\Bar{\rho}$ calculated over all $i \ne j$.

The convergence of Algorithm \ref{Algo:ProjectedGradient} is analyzed by evaluating the rate at which the objective function approaches its optimal value, measured using the metric
\begin{equation}
\varepsilon_f(k) = \frac{ f(\boldsymbol{\nu^\star}, \lambda^\star)-f(\boldsymbol{\nu}^k, \lambda^k) } { f(\boldsymbol{\nu^\star}, \lambda^\star) }. 
\end{equation}

In Section \ref{Subsec:CommSens}, we present the performance of the dual-functional waveform design described in Section \ref{Sec:JointWaveformDesign}. For sensing, the waveform matrix is designed to match a reference covariance matrix, e.g., under \gls{par} constraints, which can be quantified by the following cost 
\begin{equation}
    \varepsilon_{\mathrm{CM}} = \frac{ \Vert \boldsymbol{XX}^H/N_t - \boldsymbol{R}_0\Vert_F^2}{\Vert \boldsymbol{R}_0\Vert_F^2},
\end{equation}
where $\boldsymbol{R}_0$ is the reference covariance matrix, designed as described in Section \ref{Sec:BeampatternSynthesis}. 
For communication, we adopt \gls{ser} as the performance evaluation metric. 

\begin{figure*}%
    \centering
    \begin{subfigure}[t]{\figWidth cm}
        \includegraphics[width=\figWidth cm]{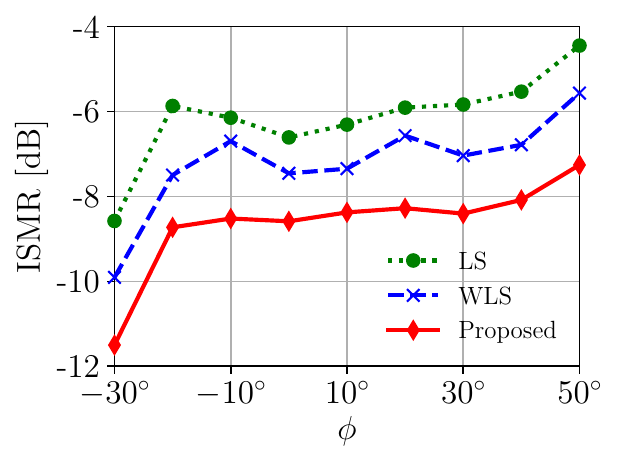}
        \subcaption{Integrated sidelobe to mainlobe ratios.}
        \label{ismr_changingdoa}
    \end{subfigure}%
    \begin{subfigure}[t]{\figWidth cm}
        \includegraphics[width=\figWidth cm]{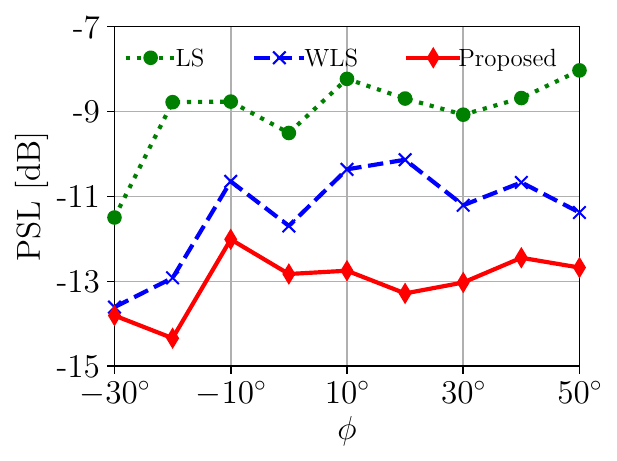}
        \subcaption{Peak sidelobe levels.}
        \label{psl_changingdoa}
    \end{subfigure}%
    \begin{subfigure}[t]{\figWidth cm}
        \includegraphics[width=\figWidth cm]{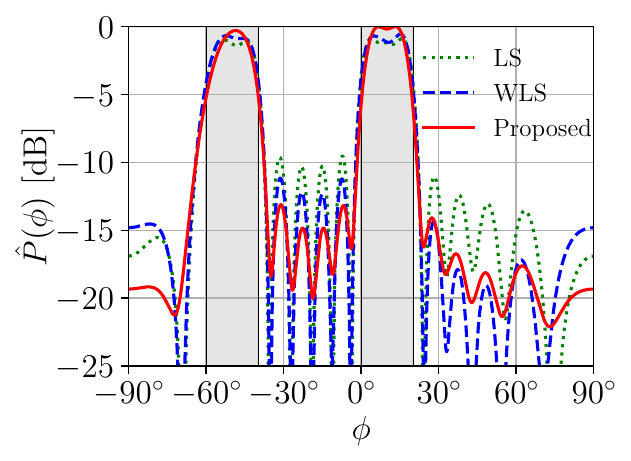}
        \subcaption{Synthesized beampattern (mainlobes at $\phi = 10^{\circ}$ and $\phi = -50^{\circ}$).}
        \label{bp_changing_doa}
    \end{subfigure}%
    \caption{Synthesized beampattern quality as a function of the scanning mainlobe direction $\phi$, with an additional fixed mainlobe at $-50^{\circ}$; the desired beamwidth for both mainlobes is $20^{\circ}$}.
\end{figure*}%

\subsection{Beampattern Synthesis Performance}
\label{Sec:BP_NumResults}
We start by evaluating the beamforming performance achieved by covariance matrices designed in the first stage of the proposed method.
The performance is compared against the most popular alternative approaches from the literature, namely the \acrlong{ls} (LS) \cite{stoica2007probing} and \acrlong{wls} (WLS) \cite{fuhrmann2008transmit} methods, which are described in Section \ref{Sec:SignalModelSensing}.
To implement the former, we use the general-purpose solver SCS \cite{o2016conic, diamond2016cvxpy}, while for the latter, we adopt the official Matlab implementation \cite{mathworksImplementation}, which is based on a standard interior-point method.  

To obtain the results presented in Sections \ref{Subsec:ChangingBW}, \ref{Subsec:ChangingDOA} and \ref{Subsec:Convergence}, we set the tuning parameter to ${\eta}=2.5 \Bar{\eta}$ in \eqref{Eqn:Tdefn}, where $\Bar{\eta}$ is defined as 
\begin{equation}
    \label{Eqn:EtaBar}
    \Bar{\eta} = P_T / \mathrm{Tr}\left(\boldsymbol{\Sigma}_D^{-2} \boldsymbol{V}_D^H \mathrm{diag}\left(\{P_d(\bar{\phi}_i)\}_{i=1}^{N_{\phi}} \right) \boldsymbol{V}_D \right).
\end{equation}
Here, setting $\eta = \Bar{\eta}$ would determine the scaling of $\boldsymbol{\Tilde{T}}$ in the unrealistic case where $\boldsymbol{\Tilde{R}} = \boldsymbol{\Tilde{T}}$ in \eqref{Eqn:ProposedOptCovMtx_simplified}.
The effect of $\eta$ on the beampattern synthesis performance and cross-correlations is investigated in Section \ref{Subsec:CrossCorr}. In all of these scenarios, we do not assume any undesired interfering source and $\boldsymbol{R}$ is constrained to satisfy $\boldsymbol{R} \in \mathcal{R}_{pa}$. In Section \ref{Subsec:SensOnly}, we examine the interference suppression performance of the proposed method where $\boldsymbol{R}$ is constrained to satisfy $\boldsymbol{R} \in \mathcal{R}_{rel}$. 

\subsubsection{Impact of Mainlobe Beamwidth}
\label{Subsec:ChangingBW}
Depending on the uncertainty of prior information regarding the radar target locations, the \gls{isac} system is desired to focus \gls{tx} power in wider or narrower regions. Therefore, the desired beamwidth can be small or large. Here we evaluate the performance of the proposed approach by changing the desired beamwidth from $10^{\circ}$ to $110^{\circ}$, with a $10^{\circ}$ step size, where the central mainlobe direction is fixed to $0^{\circ}$. 

Fig. \ref{ismr_changingbw} shows the changing \gls{ismr} as a function of the beamwidth. We observe that the proposed method allocates the available power budget more efficiently than the alternative approaches for all considered beamwidths. Compared to \gls{wls} and \gls{ls}, the proposed method obtains 1.5 dB and 2.3 dB lower \gls{ismr} on average, respectively. 
Fig. \ref{psl_changingbw} shows the changing \gls{psl} as a function of the beamwidth. The proposed method also attains lower \gls{psl} compared to the alternative approaches where it obtains 3.1 dB and 1.9 dB lower \gls{psl} on average compared to \gls{ls} and \gls{wls}, respectively.
Fig. \ref{bp_changing_bw} shows a typical synthesized beampattern obtained for a desired beamwidth of $60^{\circ}$, with the gray area indicating the desired mainlobe region. As observed, the proposed method effectively conforms to the desired mainlobe region, illustrating its suitability for synthesizing beampatterns.

\subsubsection{Impact of the Mainlobe Direction in Two-Mainlobe Case}
\label{Subsec:ChangingDOA}
A practical requirement for a beampattern synthesis algorithm is the ability to support multiple mainlobes. For instance, a radar system may need to illuminate a known target while simultaneously scanning other regions for unidentified ones.
This section evaluates the performance of the proposed beampattern synthesis approach in a scenario featuring one fixed mainlobe at $\phi=-50^\circ$ and another scanning from $\phi=-30^\circ$ to $\phi=50^\circ$ in $10^\circ$ steps. The desired beamwidth for both mainlobes is set to $20^\circ$.

Fig. \ref{ismr_changingdoa} illustrates the \gls{ismr} as a function of the scanning mainlobe's direction. The proposed method demonstrates superior power allocation to the mainlobes compared to the alternatives, outperforming \gls{wls} and \gls{ls} by an average of $1.4$ dB and $2.5$ dB, respectively. 
The \gls{psl} obtained for this scenario is presented in Fig. \ref{psl_changingdoa}. The proposed method achieves the lowest \gls{psl} values among the approaches considered, with average reductions of $1.6$ dB and $4$ dB compared to the \gls{wls} and \gls{ls} methods, respectively.
Fig. \ref{bp_changing_doa} shows a typical synthesized beampattern for the case where the scanning mainlobe is fixed at $10^\circ$. The proposed method also effectively allocates nearly equal \gls{tx} power to both mainlobes, despite their significant angular separation, demonstrating its capability to evenly illuminate two distinct target regions.

\subsubsection{Convergence and Computation Time of Algorithm \ref{Algo:ProjectedGradient}}
\label{Subsec:Convergence}
\begin{figure}
    \centering
    \includegraphics[width=7 cm]{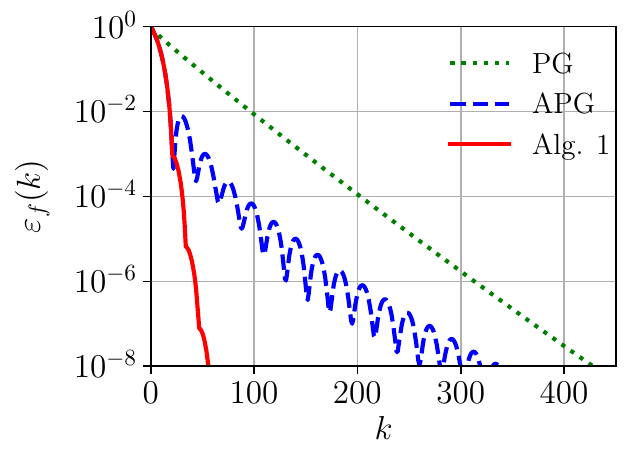}
    \caption{Convergence of Algorithm \ref{Algo:ProjectedGradient}, with $k$ denoting the iteration counter. Standard and accelerated projected gradient methods are denoted by PG and APG, respectively.}
    \label{convergence}
\end{figure}

For the simulation scenario shown in Fig. \ref{bp_changing_bw}, we provide the convergence plot of Algorithm \ref{Algo:ProjectedGradient} in Fig. \ref{convergence}.
We observe that Algorithm \ref{Algo:ProjectedGradient} demonstrates much faster convergence compared to standard (constant step-size with $\gamma=1/L$) and accelerated projected gradient methods, achieving $\varepsilon_f(k) < 10^{-8}$ after $k=50$ iterations, while the same accuracy can be obtained after $k=400$ and $k=300$ iterations of standard and accelerated projected gradient methods, respectively. 

For the scenarios described in Section \ref{Subsec:ChangingDOA}, the average beampattern synthesis times for \gls{wls}, \gls{ls}, and the proposed method (Algorithm \ref{Algo:ProjectedGradient}) are 38.8, 12.3, and 0.03 s, respectively, as measured on a computer with Intel i7-1365U CPU and 16 GB RAM.
When $M_T$ is increased from 16 to 64, for the same scenario, the \gls{ls} solution requires over 7 hours, whereas the proposed method completes in just 1.3 s on the same hardware. 

\add{
The proposed formulation \eqref{Eqn:ProposedOptCovMtx_simplified} can be also solved using a general-purpose interior-point solver. To this end, we utilize CVXPY ~\cite{diamond2016cvxpy} to obtain a solution with the modern interior-point solver Clarabel~\cite{Clarabel_2024}, under the same scenario and hardware conditions. For the case of $M_T = 64$, the results show that Clarabel requires 82.8 s to converge, i.e., substantially higher than the 1.3 s achieved by the proposed method. When $M_T = 128$, the proposed algorithm completes in approximately $13.5$ s, whereas Clarabel fails to produce a solution due to excessive memory usage. These results demonstrate the computational efficiency of the proposed method, while also highlighting its practicality for massive \gls{mimo} systems. 
}

\subsubsection{Impact of $\eta$ on Cross-Correlation and Beampattern}
\label{Subsec:CrossCorr}
\begin{figure}
    \centering
    \includegraphics[width=7.5 cm]{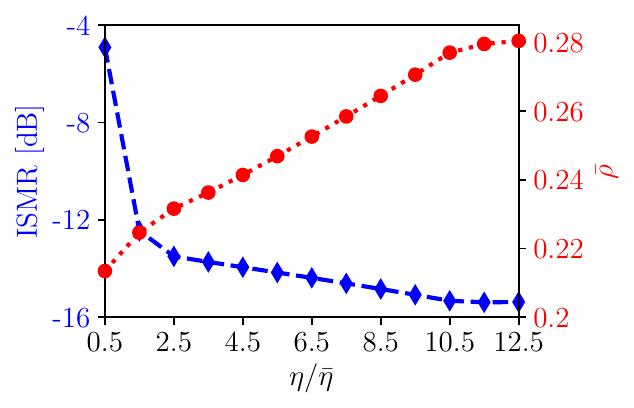}
    \caption{The effect of $\eta$ on the synthesized beampattern quality and cross-correlations.}
    \label{crosscorr_eta}
\end{figure}

The effect of the tuning parameter $\eta$ in \eqref{Eqn:Tdefn} on the cross-correlation and beampattern synthesis performnance is examined according to \eqref{Eq:Crosscorr}, by considering the scenario from Fig. \ref{bp_changing_bw}. Fig. \ref{crosscorr_eta} demonstrates that $\eta$ governs the trade-off between low cross-correlation and \gls{ismr}. As $\eta$ increases, \gls{ismr} decreases, but this comes at the expense of higher cross-correlation values.
As a reference, the average cross-correlation values ($\Bar{\rho}$) achieved by \gls{wls} and \gls{ls} are about $0.39$ in this scenario. With $\eta=2.5\Bar{\eta}$ used in the previous scenarios, the proposed method achieves a lower average cross-correlation value of $\Bar{\rho}=0.23$. 

\subsection{Communication and Sensing Performance}
\label{Subsec:CommSens}
In this section, we investigate the performance of the proposed dual-functional waveform design described in Section \ref{Sec:JointWaveformDesign}. 
In Section \ref{Subsec:CommSensTradeoff}, we demonstrate effect of the trade-off parameter $\alpha$ in \eqref{Eqn:ProposedJointDesign} on communication and sensing performance. In Sections \ref{Subsec:SensOnly} and \ref{Subsec:CommOnly}, we set $\alpha = 0$ and $\alpha = 1$, to observe the sensing- and communication-only cases, respectively. 

For the communication channel, we assume a Rayleigh fading channel, where the \gls{snr} for each user is the same and defined as $\mathrm{SNR}_i = \beta_i P_T / \sigma_c^2$, $1 \le i \le K_C$. We assume that the communication symbols are drawn from the \gls{qpsk} alphabet. The block length is set to $N_t=2048$. 

\subsubsection{Communication-Sensing Trade-off}
\label{Subsec:CommSensTradeoff}
\begin{figure}
    \centering
    \includegraphics[width=8 cm]{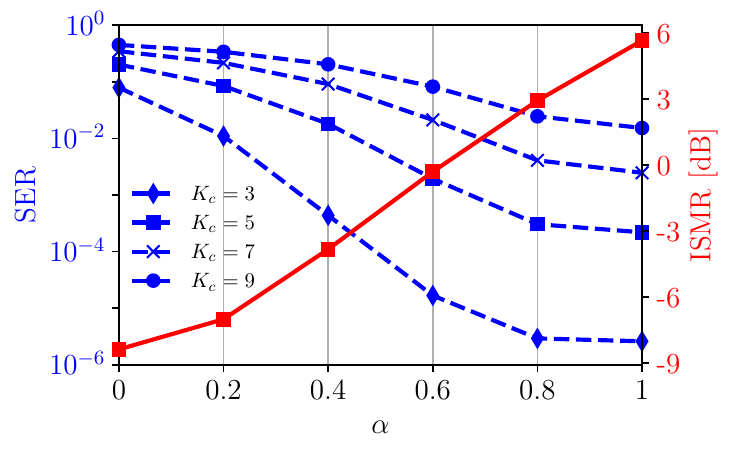}
    \caption{The effect of the tuning parameter $\alpha$ on \gls{ser} and beampattern synthesis performance, for \gls{snr} of 10 dB.}
    \label{alpha_effect}
\end{figure}

Fig. \ref{alpha_effect} illustrates the impact of the tuning parameter $\alpha$ on sensing and communication performance. The results are obtained by averaging over 1000 independent channel realizations. The sensing scenario is based on the setup depicted in Fig. \ref{bp_changing_doa}, where the waveform is constrained to satisfy the per-antenna power constraint. 
As expected, increasing $\alpha$ reduces the \gls{ser}, indicating improved communication performance. However, this comes at the cost of higher \gls{ismr} values, signifying a decline in sensing performance.
Additionally, the results reveal that increasing the number of communication users degrades the communication performance. For instance, in the high \gls{snr} regime, a \gls{ser} of $1.5 \times 10^{-2}$ is observed for $K_C = 9$, compared to $2.5 \times 10^{-6}$ for $K_C = 3$. 
\add{
We only report \gls{ismr} results for $K_C = 5$, as the \gls{ismr} remains largely unaffected by changes in $K_C$ for this specific scenario.}

\subsubsection{Sensing-Only Performance}
\label{Subsec:SensOnly}
\begin{figure}[t]
    \centering
    \includegraphics[width=7 cm]{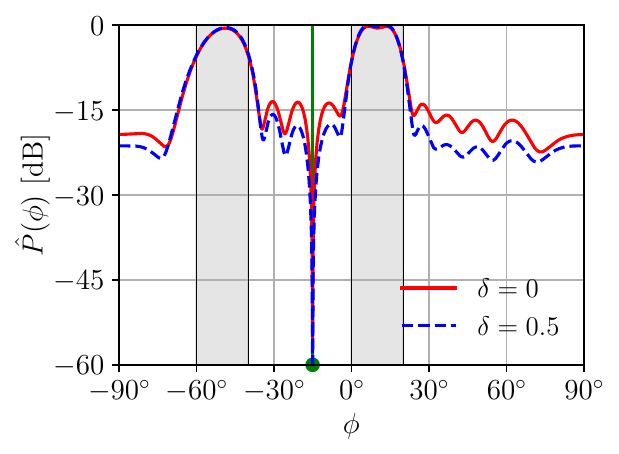}
    \caption{The synthesized beampatterns under the constraints $\boldsymbol{R}\in \mathcal{R}_{rel}$ for $\delta = 0.5$ and $\boldsymbol{R}\in \mathcal{R}_{pa}$ ($\delta = 0$) with a notch direction at $\phi = -15^{\circ}$.}
    \label{bp_relprm}
\end{figure}

For the scenario depicted in Fig. \ref{bp_changing_doa}, we consider the presence of an interference source, e.g., a jammer located at $\phi = -15^{\circ}$.
By setting $\omega = 10^{4}\eta$ in \eqref{Eqn:OveralCost1}, interference suppression of up to $-60$ dB can be achieved, as demonstrated by a typical solution shown in Fig. \ref{bp_relprm}.
Furthermore, improved beampattern synthesis performance is observed by relaxing the per-antenna power constraint with $\delta=0.5$ ($\boldsymbol{R} \in \mathcal{R}_{rel}$, see Section \ref{Subsec:HardwareImp}), which is taken as the reference covariance matrix input $\boldsymbol{R}_0$ to synthesize $\boldsymbol{X}$ in the next step. 

\begin{figure}[t]
    \centering
    \includegraphics[width=7 cm]{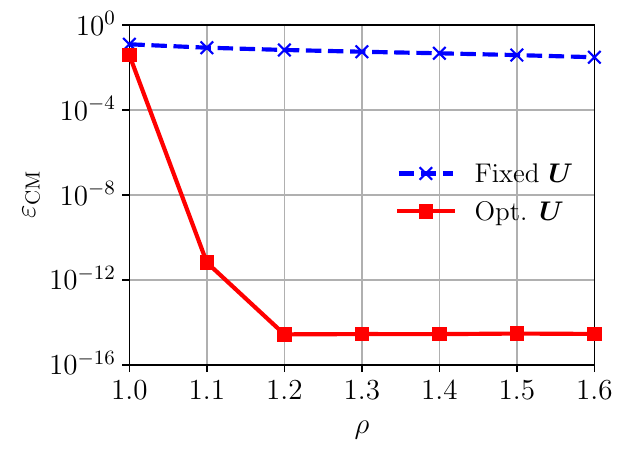}
    \caption{Covariance matrix matching performance of the sensing waveform depending on the imposed \gls{par} constraints.}
    \label{papr_sensing}
\end{figure}

\begin{figure}[t]
    \centering
    \includegraphics[width=7 cm]{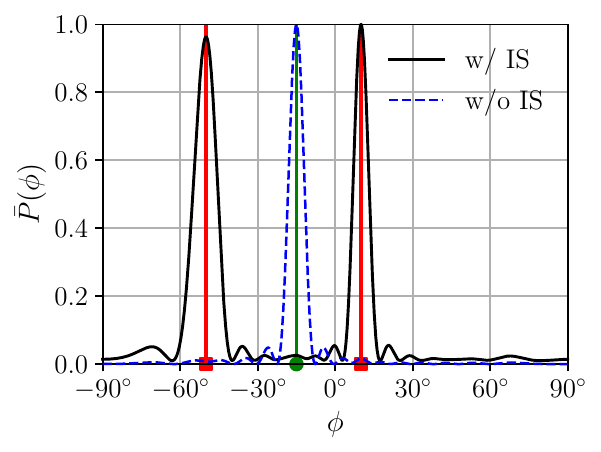}
    \caption{Normalized angle estimation spectrums with and without interference suppression (IS).}
    \label{doa_spectrum}
\end{figure}

We begin by evaluating the effectiveness of $J_{\mathrm{syn}}^s(\boldsymbol{X}, \boldsymbol{U})$ \eqref{Eqn:WaveformCostSensing} as a sensing cost. For that purpose, we design $\boldsymbol{X}$ under various \gls{par} constraints. 
For comparison, we utilize $J_{\mathrm{ref}}^s(\boldsymbol{X})$ \eqref{Eqn:SimilarityMetric} with $\boldsymbol{X}_0 = \sqrt{N_t}\boldsymbol{R}_0^{1/2} \boldsymbol{U}_0$, where the columns of the semi-unitary matrix $\boldsymbol{U}_0$ are the right singular vectors of a Gaussian random $M_T \times N_t$ matrix, with i.i.d. entries drawn from $\mathcal{CN}({0}, 1)$.
This serves as a special case of $J_{\mathrm{syn}}^s(\boldsymbol{X}, \boldsymbol{U})$ \eqref{Eqn:WaveformCostSensing} with a fixed $\boldsymbol{U}=\boldsymbol{U}_0$. %

Fig. \ref{papr_sensing} shows that optimizing over $\boldsymbol{U}$ yields significant performance enhancements. For instance, when $\rho = 1.6$, $\varepsilon_{\mathrm{CM}}$ is on the order of $10^{-15}$ when $\boldsymbol{U}$ is treated as an optimization variable, compared to $10^{-2}$ when it is fixed.
As higher \gls{par} is allowed, the covariance matching performance of the generated waveform improves. This result is expected, as lower \gls{par} imposes a stricter constraint. Furthermore, the substantial difference between the performance under $\rho = 1$ (\gls{cm}) and $\rho = 1.1$ highlights that allowing 10\% deviations from the mean power level yields significantly better performance.\looseness=-1

Fig. \ref{doa_spectrum} presents the normalized angle estimation spectrum generated using the Bartlett beamforming algorithm for an \gls{snr} of 10 dB \cite{van1988beamforming}. Here $\boldsymbol{X}$ is designed under the \gls{par} constraint of $\rho=1.1$, where the reference covariance matrix is designed
with and without considering the interference source\footnote{Neglecting the interference source corresponds to omitting $J^s_{\mathrm{is}}(\boldsymbol{R})$ term in \eqref{Eqn:OveralCost1} by setting $\omega=0$ in the covariance matrix ($\boldsymbol{R}_0$) design stage.} in Fig. \ref{bp_relprm}. The interference direction and the radar target directions are denoted by green and red lines, respectively. 

In this scenario, radar targets are modeled as $q(\phi_1) = e^{j\psi_1}$ and $q(\phi_2) = e^{j\psi_2}$, located within mainlobe directions $\phi_1 = -50^\circ$ and $\phi_2 = 10^\circ$, respectively. An interference source is represented by $q(\phi_3) = 100e^{j\psi_3}$ at $\phi_3 = -15^\circ$ (the \acrlong{sir} is $-40$ dB). Random phase shift is incorporated into the model where $\psi_i \sim U[0,2\pi]$. 
When the known interference direction is suppressed by nulling, angle estimation can be performed reliably.
Conversely, if suppression is not performed, the radar targets become indistinguishable.

\add{More advanced techniques, such as subspace-based and sparse methods \cite{krim1996two,kilic2022adaptive}, can also be employed in \gls{rx} processing to improve performance. However, a detailed discussion of these techniques is beyond the scope of this work.}

\subsubsection{Communication-Only Performance}
\label{Subsec:CommOnly}
\begin{figure}[t]
    \centering
    \includegraphics[width=7 cm]{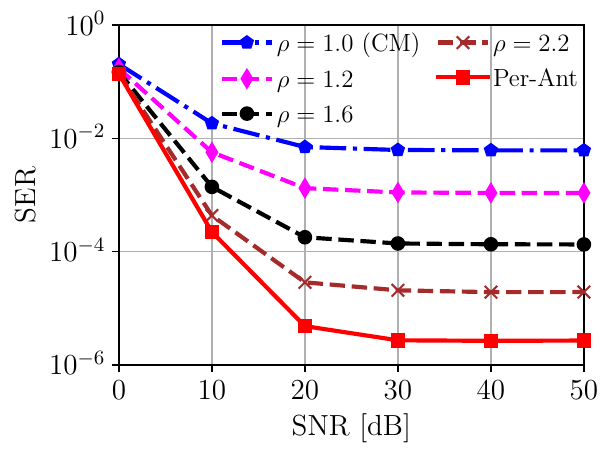}
    \caption{\gls{ser} vs \gls{snr} for per-antenna power constraint, different constraints $\rho$ on \gls{par}, and the \gls{cm} constraint, for $\alpha=1$.}
    \label{ser_vs_snr}
\end{figure}

Fig. \ref{ser_vs_snr} illustrates \gls{ser} as a function of \gls{snr}, under different hardware implementation constraints.
The presented results are obtained by averaging over 2500 independent channel realizations and for $K_C=5$.

All \gls{ser} curves display an error floor for \gls{snr} values exceeding 30 dB, with the error floor levels determined by the imposed constraints. This residual \gls{ser} arises from waveform distortion relative to the nominal waveform.
The \gls{cm} constraint proves to be the most restrictive one, leading to an error floor on the order of $10^{-2}$, while the per-antenna constraint results in a significantly lower \gls{ser} of $2.5 \times 10^{-6}$.

When \gls{par} is limited to 2.2, only a minor performance degradation is observed compared to the per-antenna power constraint. However, further reducing the \gls{par} significantly deteriorates communication performance.
This result aligns with the findings presented in Section \ref{Subsec:SensOnly} and highlights that the adjustable \gls{par} enables a flexible trade-off between hardware complexity and performance optimization.

\section{Conclusions}
\label{Sec:Conclusions}
This paper presents a flexible and computationally efficient joint waveform design framework for \gls{mimo} \gls{isac} systems, accommodating arbitrary antenna array responses and supporting various important practical constraints.
The waveform is synthesized in two main steps.  

The first one involves design of the \gls{tx} covariance matrix, formulated as a convex matrix nearness problem, incorporating the desired radiation pattern and spatial cross-correlation characteristics for the improved sensing performance. The proposed method enables independent power control over individual antennas and supports enforcement of deep radiation pattern notches at specified directions for interference suppression.
While ensuring global convergence, it is demonstrated that a highly-efficient solver can be implemented by leveraging the structure of the proposed problem formulation. 

The second step employs a waveform synthesis method to obtain \gls{tx} signal samples conforming to the covariance matrix produced in the previous step.
The proposed method employs a \gls{zf}-based \gls{mumimo} approach to minimize the interference between communication users, while ensuring compliance with practical hardware constraints.
Most notably, it enables incorporation of adjustable \gls{par} requirements individually on each antenna. 
Extensive simulation results presented in the paper validate the proposed framework's ability to deliver superior flexibility, computational efficiency, and performance compared to the existing methods in the literature.

\appendices
\section{}
\label{Sec:AppendixProgramEquality}
To simplify \eqref{Eqn:ProposedOptCovMtx2}, we rewrite $\Vert \boldsymbol{D}^H\boldsymbol{R}\boldsymbol{D} - \boldsymbol{T} \Vert_F^2$ as 
\begin{align}
    &\Vert \boldsymbol{D}^H\boldsymbol{R}\boldsymbol{D} - \boldsymbol{T} \Vert_F^2 \nonumber \\ 
    &=\mathrm{Tr}( \boldsymbol{D}^H\boldsymbol{R}\boldsymbol{D} \boldsymbol{D}^H\boldsymbol{R}\boldsymbol{D}) 
    - 2\mathrm{Re}\{\mathrm{Tr}(\boldsymbol{D}^H\boldsymbol{R}\boldsymbol{DT})\} + \Vert \boldsymbol{T} \Vert_F^2.  
\end{align}
Note that 
\begin{align}
    &\mathrm{Tr}( \boldsymbol{D}^H\boldsymbol{R}\boldsymbol{D} \boldsymbol{D}^H\boldsymbol{R}\boldsymbol{D}) \nonumber \\
    &= \mathrm{Tr}( (\boldsymbol{U}_D\boldsymbol{\Sigma}_D)^H\boldsymbol{R}(\boldsymbol{U}_D\boldsymbol{\Sigma}_D) (\boldsymbol{U}_D\boldsymbol{\Sigma}_D)^H\boldsymbol{R}(\boldsymbol{U}_D\boldsymbol{\Sigma}_D)),
\end{align}
and
\begin{equation}
    \mathrm{Tr}( \boldsymbol{D}^H\boldsymbol{R}\boldsymbol{DT} ) = \mathrm{Tr}((\boldsymbol{U}_D\boldsymbol{\Sigma}_D)^H\boldsymbol{R}(\boldsymbol{U}_D\boldsymbol{\Sigma}_D)\boldsymbol{V}_D^H\boldsymbol{T}\boldsymbol{V}_D).
\end{equation}
By \eqref{Eqn:RtildeDefn}, we can express
\begin{equation}
    \min_{\boldsymbol{R}} \Vert \boldsymbol{D}^H\boldsymbol{R}\boldsymbol{D} - \boldsymbol{T} \Vert_F^2 \equiv \min_{\boldsymbol{\tilde{R}}} \Vert \boldsymbol{\tilde{R}} - \boldsymbol{V}_D^H\boldsymbol{T}\boldsymbol{V}_D\Vert_F^2,
\end{equation}
since  
\begin{align}
    \mathrm{Tr}(\boldsymbol{\tilde{R}}^2) - 2\mathrm{Re}\{ \mathrm{Tr}(\boldsymbol{\tilde{R}}\boldsymbol{V}_D^H\boldsymbol{T}\boldsymbol{V}_D) \} + \Vert \boldsymbol{T} \Vert_F^2 \nonumber \\
    = \Vert \boldsymbol{\tilde{R}} - \boldsymbol{V}_D^H\boldsymbol{T}\boldsymbol{V}_D\Vert_F^2 + \mathrm{const},
\end{align}
where the term $\mathrm{const}$ does not depend on $\Tilde{\boldsymbol{R}}$. We can then modify the affine terms in \eqref{Eqn:ProposedOptCovMtx2} using the definition of $\boldsymbol{\tilde{R}}$ given in \eqref{Eqn:RtildeDefn}. 
Also, $\boldsymbol{\tilde{R}}\succeq \boldsymbol{0}$ if and only if $\boldsymbol{R}\succeq \boldsymbol{0}$. Hence, \eqref{Eqn:ProposedOptCovMtx_simplified} is obtained. 

\section{}
\label{Sec:AppendixDualDerivation}
Let $\nu_i$'s, $\lambda$ and $\boldsymbol{G}\succeq \boldsymbol{0}$ be the Lagrange multipliers associated with the affine inequality constraints, the affine equality constraint and the positive semidefiniteness constraint, respectively. By the definitions of $\boldsymbol{\Bar{E}}_i$ and $\boldsymbol{b}$ given in \eqref{Eqn:DefinitionsDual1} and {\eqref{Eqn:DefinitionsDual3}}, respectively, the Lagrangian associated with \eqref{Eqn:ProposedOptCovMtx_simplified} can be written as
\begin{align}
\label{Eqn:Lagrange2}
    L(\boldsymbol{{\Tilde{R}}}, \boldsymbol{G}, \boldsymbol{\nu}, \lambda) = \frac{1}{2} \Vert \boldsymbol{{\Tilde{R}}} -\boldsymbol{{\Tilde{T}}}\Vert_F^2 - \boldsymbol{\nu}^T\boldsymbol{b}-\lambda P_T \nonumber \\ + \mathrm{Tr}\left(\boldsymbol{\Tilde{R}}\left(-\boldsymbol{G}+\boldsymbol{\Psi}+ \sum_{i=1}^{2M_T} \nu_i \boldsymbol{\Bar{E}}_i + \lambda \boldsymbol{C}   \right) \right).
\end{align}
We set the gradient of $L(\boldsymbol{{\Tilde{R}}}, \boldsymbol{G}, \boldsymbol{\nu}, \lambda)$ with respect to $\boldsymbol{\Tilde{R}}$ to zero to minimize $L(\boldsymbol{{\Tilde{R}}}, \boldsymbol{G}, \boldsymbol{\nu}, \lambda)$ over $\boldsymbol{\Tilde{R}}$, i.e., 
\begin{equation}
    \nabla_{\boldsymbol{\Tilde{R}}} L(\boldsymbol{{\Tilde{R}}}, \boldsymbol{G}, \boldsymbol{\nu}, \lambda) = 0,
\end{equation}
which gives  
\begin{equation}
    \label{Eqn:ArgMinR}
    \boldsymbol{\tilde{R}}(\boldsymbol{G},\boldsymbol{\nu},\lambda) =  \boldsymbol{\Tilde{T}} + \boldsymbol{G} - \boldsymbol{\Psi} - \boldsymbol{E}(\boldsymbol{\nu}) - \boldsymbol{C}(\lambda),  
\end{equation}
where $\boldsymbol{E}(\boldsymbol{\nu})$ and $\boldsymbol{C}(\lambda)$ are defined as in \eqref{Eqn:DefinitionsDual2}.  

Here we adopt the dual method proposed in \cite{boyd2005least}.  
Substituting \eqref{Eqn:ArgMinR} in \eqref{Eqn:Lagrange2} gives
\begin{align}
    g(\boldsymbol{G}, \boldsymbol{\nu}, \lambda) = \frac{1}{2} \Vert \boldsymbol{G} - \boldsymbol{\Psi} - \boldsymbol{E}(\boldsymbol{\nu}) - \boldsymbol{C}(\lambda) \Vert_F^2- \boldsymbol{\nu}^T&\boldsymbol{b}- \lambda P_T \nonumber \\ 
    + \mathrm{Tr}\big( (\boldsymbol{\Tilde{T}} + \boldsymbol{G} - \boldsymbol{\Psi} - \boldsymbol{E}(\boldsymbol{\nu}) - \boldsymbol{C}(\lambda)) & \nonumber \\ (-\boldsymbol{G}+\boldsymbol{\Psi}+ \boldsymbol{E}(\boldsymbol{\nu})+ \boldsymbol{C}(\lambda) &  ) \big),
\end{align}
which can be simplified as 
\begin{align}
    \label{Eqn:g}
    g(\boldsymbol{G}, \boldsymbol{\nu}, \lambda) = -\frac{1}{2} \Vert \boldsymbol{\Tilde{T}} + \boldsymbol{G} - \boldsymbol{\Psi} - &\boldsymbol{E}(\boldsymbol{\nu}) - \boldsymbol{C}(\lambda) \Vert_F^2 \nonumber \\ + \frac{1}{2}\Vert & \boldsymbol{\Tilde{T}} \Vert_F^2 -\boldsymbol{\nu}^T\boldsymbol{b} - \lambda P_T.  
\end{align}
The dual problem of \eqref{Eqn:ProposedOptCovMtx_simplified} can then be written as
\begin{equation}
    \label{Eqn:DualProb}
    \max_{\boldsymbol{G}, \boldsymbol{\nu}, \lambda} g(\boldsymbol{G}, \boldsymbol{\nu}, \lambda) ~~\mathrm{s.t.}~~ \boldsymbol{G}\succeq \boldsymbol{0}, ~\boldsymbol{\nu}\ge \boldsymbol{0}. 
\end{equation}
We can analytically solve \eqref{Eqn:DualProb} over $\boldsymbol{G}$ by \cite{higham2002computing}
\begin{equation}
    \label{Eqn:GMaximizer}
    \boldsymbol{G}= \big[\boldsymbol{\Tilde{T}} - \boldsymbol{E}(\boldsymbol{\nu}) - \boldsymbol{C}(\lambda)- \boldsymbol{\Psi} \big]_-,
\end{equation}
where 
\begin{equation}
    \label{Eqn:MEquality}
    [\boldsymbol{M}]_- =   [\boldsymbol{M}]_+ - \boldsymbol{M}. 
\end{equation}
Then, by placing \eqref{Eqn:GMaximizer} into \eqref{Eqn:g}, we obtain 
\begin{equation}
    \label{Eqn:DualObj_}
    f(\boldsymbol{\nu}, \lambda) = -g\big(\big[\boldsymbol{\Tilde{T}} - \boldsymbol{E}(\boldsymbol{\nu}) - \boldsymbol{C}(\lambda)-\boldsymbol{\Psi} \big]_-, \boldsymbol{\nu}, \lambda \big),
\end{equation}
and by placing \eqref{Eqn:GMaximizer} into \eqref{Eqn:ArgMinR} and using \eqref{Eqn:MEquality}, we simplify \eqref{Eqn:DualObj_} as \eqref{Eqn:DualObj} and obtain \eqref{Eqn:SimplifiedDualProb}.

\bibliographystyle{IEEEtran}
\bibliography{utils/journal_refs}

\end{document}